\documentstyle[psfig]{ccsr}         

\begin{document}

% To make footnotesize text-style subscripts in math formulas
\newcommand{\textsub}[1]
{\mbox{\footnotesize #1}}

% To make tiny text-style subscripts in math formulas
\newcommand{\tinytextsub}[1]
{\mbox{\tiny #1}}

% To make bibliography entries
\newcommand{\bibentry}[1]
{\hangindent=.5in
 \hangafter=1
 \noindent #1}

% Notation for CAs

% CA with numerical subscript
\newcommand{\ca}[1]
{$\phi_{#1}$}

% CA with text subscript 
\newcommand{\caT}[1]
{$\phi_{\textsub{#1}}$}

% CA with text subscript and superscript
\newcommand{\caS}[2]
{$\phi^{\textsub{#2}}_{\textsub{#1}}$}

%\newcommand{\parA}[0]
%{\caS{par}{a}}

%\newcommand{\parB}[0]
%{\caS{par}{b}}

% Notation for fitness
\newcommand{\fitness}[3]
{$F^{#1}_{#2}$(#3)}

% Fitness, no parens
\newcommand{\fit}[2]
{$F^{#1}_{#2}$}

% Notation for performance
\newcommand{\performance}[3]
{${\cal P}^{#1}_{#2}$(#3)}

% Performance, no parens
\newcommand{\perf}[2]
{${\cal P}^{#1}_{#2}$}

% Abbreviation for figure references
\newcommand{\fig}[1]
{Fig.~\ref{#1}}

% Abbreviation for plural figure references 
\newcommand{\figs}[1]
{Figures~\ref{#1}}

% Notation for regular domains
\newcommand{\domain}[1]
{$\Lambda^{#1}$}

\begin{center}
{\large\bf The Evolutionary Design of Collective Computation\\
in \\
Cellular Automata}

\parbox{2.1in}
{\begin{center}
James P. Crutchfield$^\dag$\\
Santa Fe Institute \\
1399 Hyde Park Road\\
Santa Fe, NM 87501 \\
chaos@santafe.edu
\end{center}}
\parbox{2.1in}
{\begin{center}
Melanie Mitchell \\
Santa Fe Institute \\
1399 Hyde Park Road\\
Santa Fe, NM 87501 \\
mm@santafe.edu
\end{center}}
\parbox{2.1in}
{\begin{center}
Rajarshi Das$^\ddag$ \\
Center for Nonlinear Studies \\
Los Alamos Nat'l Laboratory \\
Los Alamos, NM 87545 \\
raja@cnls.lanl.gov
\end{center}}

$^\dag$also Physics Department, University of California,
Berkeley, CA 94720\\
$^\ddag$also IBM Watson Research Center, P.O. Box 704, 
Yorktown Heights, NY 10598 
\end{center}

\begin{abstract} 
  We investigate the ability of a genetic algorithm to design cellular
  automata that perform computations. The computational strategies of
  the resulting cellular automata can be understood using a framework
  in which ``particles'' embedded in space-time configurations carry
  information and interactions between particles effect information
  processing.  This structural analysis can also be used to explain the
  evolutionary process by which the strategies were designed by the
  genetic algorithm. More generally, our goals are to understand how
  machine-learning processes can design complex decentralized systems
  with sophisticated collective computational abilities and to develop
  rigorous frameworks for understanding how the resulting dynamical
  systems perform computation.
\end{abstract}

\tableofcontents

%%%%%%%%%%%%%%%%%%%%%%%%%% INTRODUCTION %%%%%%%%%%%%%%%%%%%%%%%%%% 

\section{Introduction}

From the earliest days of computer science, researchers have been
interested in making computers and computation more like
information-processing systems in nature. In the 1940's and 1950's, von
Neumann viewed the new field of ``automata theory'' as closely related
to theoretical biology, and asked questions such as ``How are computers
and brains alike?'' \cite{VonNeumann58} and ``What is necessary for an
automaton to reproduce itself?'' \cite{VonNeumann66}. Turing was deeply
interested in the mechanical roots of human-like intelligence
\cite{Turing50}, and Weiner looked for links among the functioning of
computers, nervous systems, and societies \cite{Weiner48}. More
recently work on biologically and sociologically inspired computation
has received renewed interest; researchers are borrowing
information-processing mechanisms found in natural systems such as
brains \cite{Churchland&Sejnowski92,Fiesler&Beale97,RumelhartEtAl86b},
immune systems \cite{FarmerEtAl86,ForrestEtAl97}, insect colonies
\cite{ColorniEtAl92,DeneubourgEtAl91}, economies
\cite{WaldspurgerEtAl92,Wellman93}, and biological evolution
\cite{Back96a,Fogel95a,Holland92a}.  The motivation behind such
work is both to understand how systems in nature adaptively process
information and to construct fast, robust, adaptive computational
systems that can learn on their own and perform well in many
environments.

Although there are some commonalities, natural systems differ
considerably from traditional von Neumann-style
architectures\footnote{It should be noted that although computer
  architectures with central control, random access memory, and serial
  processing have been termed ``von Neumann style'', von Neumann was
  also one of the inventors of ``non von Neumann-style'' architectures
  such as cellular automata.}.  Biological systems such as brains,
immune systems, and insect societies 
consist of myriad relatively homogeneous components that are extended
in space and operate in parallel with no central control and with only
limited communication among components.  Information processing in
such systems arises from coordination among large-scale patterns that
are distributed across components (e.g., distributed activations of
neurons or activities of antibodies).  Such decentralized systems,
being highly nonlinear, often exhibit complicated,
difficult-to-analyze, and unpredictable behavior. The result is that
they are hard to control and ``program''. It seems clear that in order
to design and understand decentralized systems and to develop them into
useful technologies, engineers must extend traditional notions of
computation to encompass these architectures. This has been done to
some extent in research on parallel and distributed computing (e.g.,
\cite{Crichlow97}) and with architectures such as systolic arrays
\cite{Kung82}. However, as computing systems become more parallelized
and decentralized and employ increasingly simple individual
processors, it becomes harder and harder to design and program such
systems. 

Cellular automata (CAs) are a simple class of systems that captures
some of the features of systems in nature listed above: large numbers
of homogeneous components (simple finite state machines) extended in
space, no central control, and limited communication among components.
Given that there is no programming paradigm for implementing parallel
computations in CAs, our research investigates how genetic algorithms
(GAs) can evolve CAs to perform computations requiring
coordination among many cells.  In
other words, the GA's job is to design ways in which the actions of
simple components with local information and communication give rise
to coordinated global information processing.  In addition, we have
adapted a framework---``computational mechanics''---that can be used
to discover how information processing is embedded in dynamical
systems \cite{Crut92c} and thus to analyze how computation emerges in
evolved CAs.  Our ultimate motivations are two-fold: (i) to understand
collective computation and its evolution in natural systems and (ii)
to explore ways of automatically engineering sophisticated collective
computation in decentralized multi-processor systems.

In previous work we described some of the mechanisms by which genetic
algorithms evolve cellular automata to perform computations, and some
of the impediments faced by the GA \cite{MitchellEtAl94a}.  We also
briefly sketched our adaptation of the computational mechanics
approach to understanding computation in the evolved CAs
\cite{Crutchfield&Mitchell95a,DasEtAl95a,DasEtAl94a}.  In this paper
we give a more fully developed account of our research to date on
these topics, report on new results, and compare our work with other
work on GAs, CAs, and distributed computing.  

This paper is organized as follows.  In
Sec.~\ref{ca}--\ref{evolving-ca}, we review cellular automata, define
a computational task for CAs---``density classification''---that
requires global coordination, and describe how we used a GA to evolve
cellular automata to perform this task.  In
Sec.~\ref{results-of-experiments}--\ref{evolutionary-history}, we
describe the results of the GA evolution of CAs.  We first describe
the different types of CA computational strategies discovered by the
GA for performing the density classification task.  We then make the
notion of computational ``strategies'' more rigorous by defining them
in terms of embedded particles, particle interactions, and geometric
``subroutines'' consisting of these components.  This
high-level description enables us to explain how the space-time
configurations generated by the evolved CAs give rise to collective
computation and to predict quantitatively the CAs's computational
performance.  We then use embedded-particle descriptions to
explain the evolutionary stages by which the successful CAs were
produced by the GA.  Finally, in Sec.~\ref{related-work} we compare
our research with related work.

%%%%%%%%%%%%%%%%%%%%%%%%%% CELLULAR AUTOMATA %%%%%%%%%%%%%%%%%%%%%%%%%% 

\section{Cellular Automata \label{ca}}

%%%%%%%%%%%%%%%%%%% CA-PICTURE %%%%%%%%%%%%%%%%%%%  

\begin{figure}
\centerline{\psfig{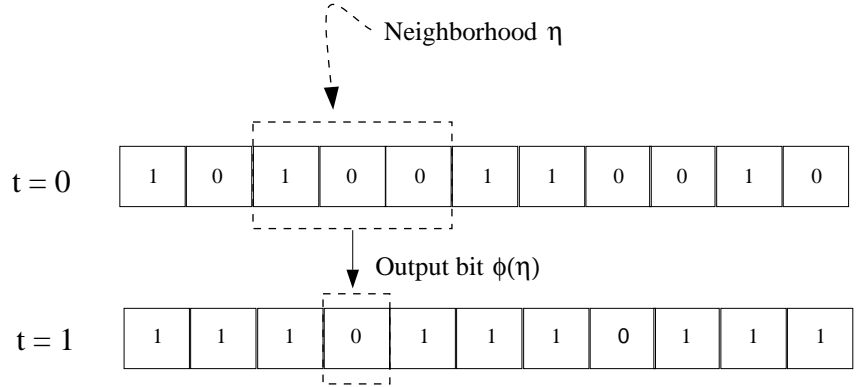}}
\caption{The components of one-dimensional, binary-state, $r=1$
  (``elementary'') CA 110 shown iterated one time step on a
  configuration with $N=11$ lattice sites and periodic boundary
  conditions (i.e., $s_{N} = s_{0}$).
\label{ca-picture}}
\end{figure}

%%%%%%%%%%%%%%%%%%%%%%%%%%%%%%%%%%%%%%%%%%%%%%%%%%%  

An one-dimensional cellular automaton consists of a lattice of $N$ identical
finite-state machines ({\it cells}), each with an identical topology
of local connections to other cells for input and output, along with
boundary conditions.  Let $\Sigma$ denote the set of states in a
cell's finite-state machine and let $k=|\Sigma|$ denote the number of
states per cell. Each cell is indexed by its site number $i =
0,1,\ldots, N-1$. A cell's state at time $t$ is denoted by $s^t_i$,
where $s^t_i \in \Sigma$.  The state $s^t_i$ of cell $i$ together with
the states of the cells to which it is connected is called the
{\it neighborhood} $\eta^t_i$ of cell $i$.  Each cell obeys the same
transition rule $\phi(\eta^t_i)$, that gives the
update state $s^{t+1}_i = \phi(\eta^t_i)$ for cell $i$ as a function
of $\eta^t_i$.  We will drop the indices on $s^t_i$ and $\eta^t_i$
when we refer to them as general (local) variables.

We use ${\bf s}^t$ to denote the {\it configuration} of cell states: 
$${\bf s}^t = s^t_0 s^t_1 \ldots s^t_{N-1}.$$
A CA $\{ \Sigma^N , \phi \}$ thus specifies a global map $\Phi$
of the configurations:
$$\Phi: {\Sigma^N} \rightarrow {\Sigma^N},$$ 
with
$${\bf s}^{t+1} = \Phi({\bf s}^t).$$
In some cases in the discussion
below, $\Phi$ will also be used to denote a map on subconfigurations
of the lattice.  Whether $\Phi$ applies to global configurations or
subconfigurations should be clear from context.

In a {\it synchronous} CA, a global clock provides an update signal for
all cells: at each $t$ all cells synchronously read the states of
the cells in their neighborhood and then update their own states
according to $s_i^t = \phi(\eta^t_i)$.

The neighborhood $\eta$ is often taken to be spatially symmetric.
For one-dimensional CAs,
$\eta_i = s_{i-r}, \ldots, s_0, \ldots, s_{i+r}$, where $r$ is
the CA's {\it radius}. Thus, $\phi: \Sigma^{2r + 1} \rightarrow \Sigma$.
For small-radius, binary-state CAs, in which the number of possible
neighborhoods is not too large, $\phi$ is often displayed as a look-up
table, or {\it rule table}, that lists each possible $\eta$ together
with its resulting {\it output bit} $s^{t+1}$.

The architecture of a one-dimensional, $(k,r) = (2,1)$ CA is illustrated
in \fig{ca-picture}. Here, the neighborhood of each cell consists of
itself and its two nearest neighbors and the boundary conditions are
periodic: $s_N = s_0$.

The 256 one-dimensional, $(k,r) = (2,1)$ CAs are called {\it elementary
CAs} (ECAs). Wolfram \cite{Wolfram83} introduced a numbering scheme for
one-dimensional CAs. The output bits can be ordered lexicographically,
as in \fig{ca-picture}, and are interpreted as the binary representation
of an integer between 0 and 255 with the leftmost bit being the
least-significant digit and the rightmost the most-significant digit.
In this scheme, the elementary CA pictured here is number $110$.

In this paper we will restrict our attention to synchronous,
one-dimensional, $(k,r)=(2,3)$ CAs with periodic boundary conditions.
This choice of parameters will be explained below. For ease of
presentation, we will sometimes refer to a CA by its transition rule
$\phi$ (e.g., as in ``the CA $\phi$ \ldots'').

%%%%%%%%%%%%%%%%%%%  ECA110 %%%%%%%%%%%%%%%%%%%  

\begin{figure}
\centerline{\psfig{figure=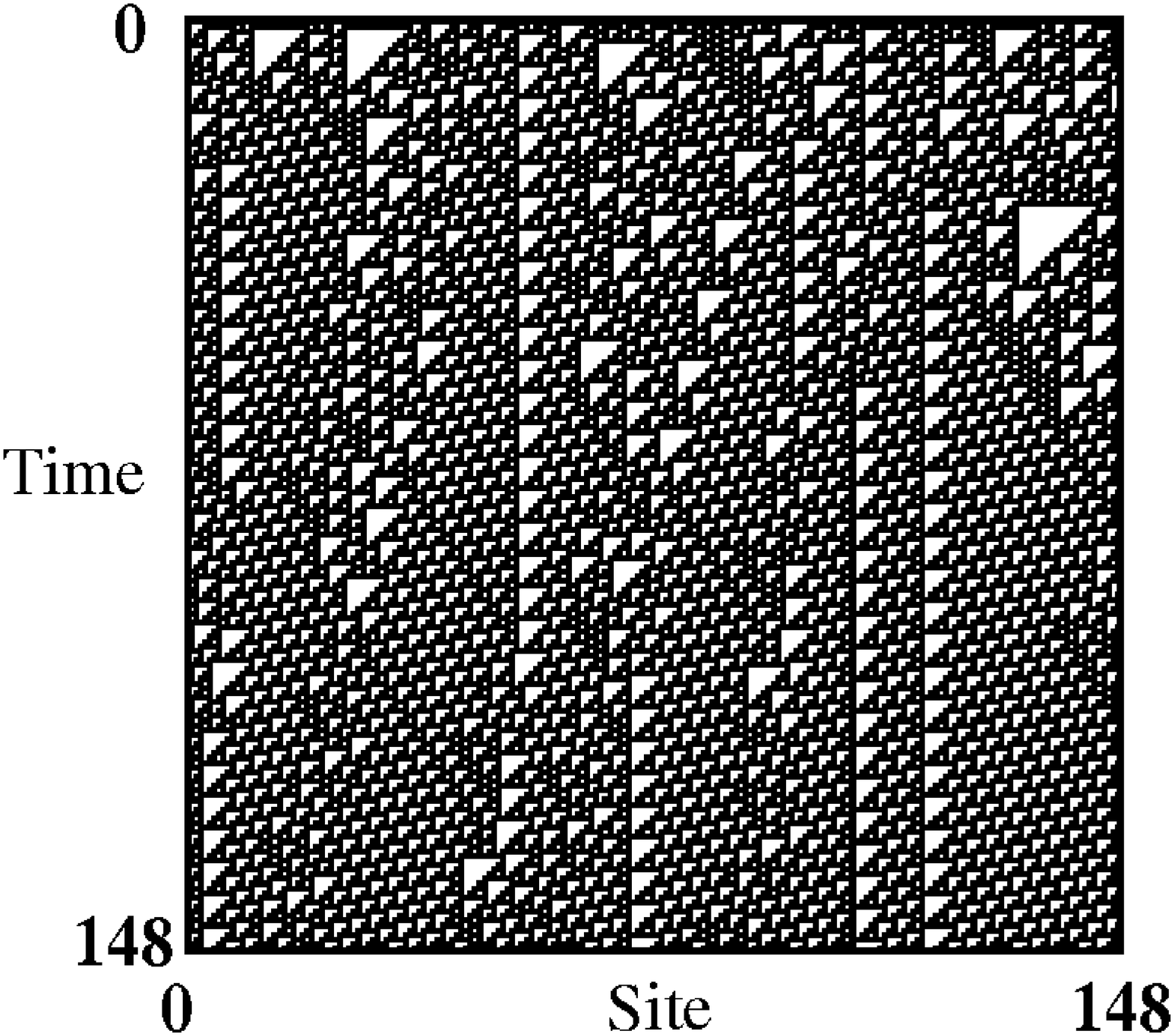,height=3in}}
\caption{A space-time diagram illustrating the typical behavior of
  elementary CA (ECA) 110. The lattice of 149 sites, displayed
  horizontally at the top, starts with ${\bf s}_0$ being an
  arbitrary initial configuration.  Cells in state 1 are displayed as
  black and cells in state 0 are displayed as white. Time increases
  down the page.
  \label{eca110}}
\end{figure}

%%%%%%%%%%%%%%%%%%%%%%%%%%%%%%%%%%%%%%%%%%%%%%%%%%%  

The behavior of CAs is often illustrated using space-time diagrams in
which the configurations ${\bf s}^t$ on the lattice are plotted as a
function of time. \fig{eca110} shows a space-time diagram of the
behavior of ECA 110 on a lattice of $N=149$ sites and periodic
boundary conditions, starting from an arbitrary initial
configuration (the lattice is displayed horizontally) and iterated
over 149 time steps with time increasing down the figure. A variety
of local structures are apparent to the eye in the space-time diagram.
They develop over time and move in space and interact.

ECAs are among the simplest spatial dynamical systems: discrete in
time, space, and local state. Despite this, as can be seen in
\fig{eca110}, they generate quite complicated, even apparently
aperiodic behavior. The architecture of a CA can be modified in many
ways---increasing the number of spatial dimensions, the number $k$ of
states per cell, and the neighborhood size $r$; modifying the boundary
conditions; making the local CA rule $\phi$ probabilistic rather than
deterministic; making the global update $\Phi$ asynchronous;
and so on.

CAs are included in the general class of ``iterative networks'' or
``automata networks''. (See \cite{FogelmanSoulieEtAl87} for a review.)
They are distinguished from other architectures in this class by their
homogeneous and local ($r \ll N$) connectivity among cells, homogeneous
update rule across all cells, and (typically) relatively small $k$.

For quite some time, due to their appealingly simple architecture, CAs
have been successfully employed as models of physical, chemical,
biological, and social phenomena, such as fluid flow, galaxy formation,
earthquakes, chemical pattern formation, biological morphogenesis,
and vehicular traffic dynamics. They have been considered
as mathematical objects about which formal properties can be proved.
They have been used as parallel computing devices, both for the
high-speed simulation of scientific models and for computational tasks
such as image processing. In addition, CAs have been used as abstract
models for studying ``emergent'' cooperative or collective behavior in
complex systems. For discussions of work in all these areas, see, e.g.,
\cite{Burks70a,FarmerEtAL84a,FogelmanSoulieEtAl87,Gutowitz90a,Kara94a,Nage96a,PARCELLA94,Mitchell98a,Toffoli&Margolus87,Wolfram86a}.

%%%%%%%%%%%%%%%%%%%% COMPUTATIONAL TASK %%%%%%%%%%%%%%%%%%%%%%%%%%%

\section{A Computational Task for Cellular Automata \label{computational-task}}

It has been shown that some CAs are capable of universal computation;
see, e.g., \cite{BerlekampEtAl82,Lindgren&Nordahl90,Smith71}. The
constructions either embed a universal Turing machine's tape states,
read/write head location, and finite-state control in a CA's
configurations and rule or they design a CA rule, supporting propagating
and interacting particles, that simulates a universal logic circuit.
These constructions are intended to be in-principle demonstrations of
the potential computational capability of CAs, rather than
implementations of practical computing devices; they do not give much
insight about the computational capabilities of CAs in practice.
Also, in such constructions it is typically very difficult to design
initial configurations that perform a desired computation.  Moreover,
these constructions amount to using a massively parallel architecture
to simulate a serial one.

Our interest in CA computation is quite different from this approach.
In our work, CAs are considered to be massively parallel and spatially
extended pattern-forming systems. Our goal is to use machine-learning
procedures, such as GA stochastic search, to automatically design CAs
that implement parallel computation by taking advantage of the
patterns formed via collective behavior of the cells.

To this end, we chose a particular computation for a one-dimensional,
binary-state CA---density classification---that requires collective
behavior.  The task is to determine whether $\rho_0$, the fraction of $1$s
in the initial configuration (IC) ${\bf s}_0$, is greater than or less than
a critical value $\rho_c$.  If $\rho_0 > \rho_c$, the entire lattice
should relax to a fixed point of all $1$s (i.e., $\Phi(1^{N}) = 1^{N})$ 
in a maximum of
$T_{\textsub{max}}$ time steps; otherwise it should relax to a fixed
point of all $0$s ((i.e., $\Phi(0^{N}) = 0^{N})$
within that time.  The task is undefined for $\rho_0
= \rho_c$.  In our experiments we set $\rho_c = 1/2$ and
$T_{\textsub{max}} = 2N$.  The {\it performance}
\performance{I}{N}{$\phi$} of a CA $\phi$ on this task is calculated
by randomly choosing $I$ initial configurations on a lattice of $N$
cells, iterating $\phi$ on each IC for a maximum of $T_{max}$ time
steps, and determining the fraction of the $I$ ICs that were correctly
classified by $\phi$---a fixed point of all $1$s for $\rho_0 >
\rho_c$, and a fixed point of all $0$s otherwise.  No partial credit
is given for final configurations that have not reached an all-$1$s or
all-$0$s fixed point.  As a shorthand, we will refer to this task as
the ``$\rho_c = 1/2$'' task.  Defining the task for other values of
$\rho_c$ is of course possible; e.g., Chau et al. showed that is is
possible to perform the task for rational densities $\rho_c$ using two
one-dimensional elementary CAs in succession \cite{ChauEtAl97}.

This task is trivial for a von Neumann-style architecture that
holds the IC as an array in memory: it simply requires counting the number
of $1$s in ${\bf s}_0$. It also trivial for a two-layer neural network
presented with each $s^0_i$ on one of its $N$ input units, all of which
feed into a single output unit: it simply requires weights set so that
the output unit fires when the activation reaches the desired threshold
$\rho_c$. In contrast, it is nontrivial to design a CA of our type to
perform this task: all cells must agree on a global characteristic of
the input even though each cell communicates its state only to its
neighbors.

The $\rho_c = 1/2$ task for CAs can be contrasted with the
well-studied tasks known as ``Byzantine agreement'' and ``consensus''
in the distributed computing literature (e.g.,
\cite{Dolev&Strong83,Farrag&Dawson88}).  These are tasks requiring a
number of distributed processors to come to agreement on a particular
value held initially by one of the processors.  Many decentralized
protocols have been developed for such tasks. They invariably assume
that the individual processors have more sophisticated computational
capabilities and memory than the individual cells in our binary-state CAs
or that the communication topologies are more complicated than that of
our CAs.  Moreover, to our knowledge, none of these protocols addresses
the problem of classifying a global property (such as initial density)
of all the processors.

Given this background, we asked whether a GA could design CAs in which
collective behavior allowed them to perform well above chance
($>$ \performance{I}{N}{$\phi$} $= 0.5$) on this task for a range of
$N$. To minimize local processor and local communication complexity,
we wanted to use the smallest values of $k$ and $r$ for which such
behavior could be obtained.  Over all 256 ECAs $\phi$, the maximum
performance \performance{10^4}{N}{$\phi$} is approximately $0.5$ for
$N \in \{149, 599, 999\}$. For all CAs $\phi$ evolved in 300 runs of
the GA on $(k,r) = (2,2)$ CAs, the maximum
\performance{10^4}{N}{$\phi$} was approximately $0.58$ for $N=149$ and
approximately $0.5$ for $N \in \{599, 999\}$. (The GA's details will
be given in the next section.)  Increasing the radius to $r=3$,
though, resulted in markedly higher performance and more sophisticated
collective behavior. As a result, all of the experiments described
in this paper were performed on one-dimensional $(k,r)=(2,3)$ CAs
with $N \in \{149, 599, 999\}$ and periodic boundary conditions.
Note that for $r=3$, the neighborhood size $|\eta| = 7$. 

%%%%%%%%%%%%%%%%%%% MAJORITY CA %%%%%%%%%%%%%%%%%%%  

\begin{figure}
\centerline{\psfig{figure=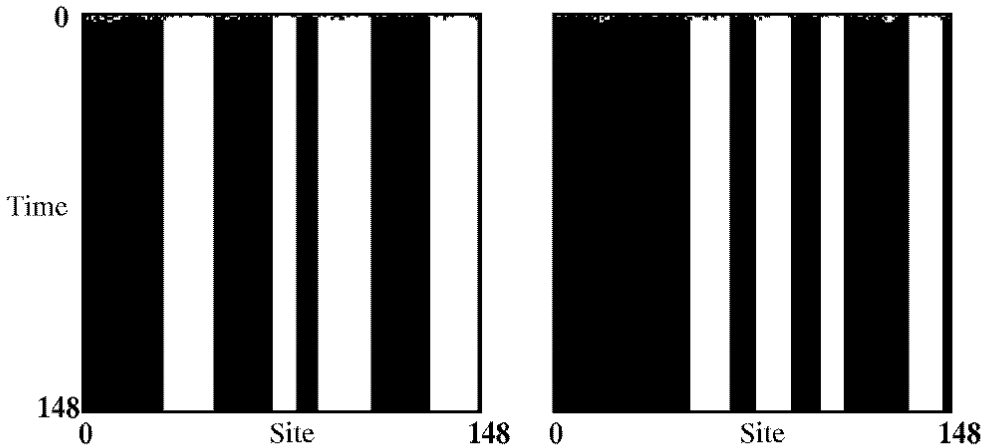,height=3in}}
\caption{Space-time diagrams for \caT{maj},  
  the $r=3$ local-majority-vote CA.  In the left diagram, $\rho_0 <
  1/2$; in the right diagram, $\rho_0 > 1/2$.}
\label{majority}
\end{figure}

%%%%%%%%%%%%%%%%%%%%%%%%%%%%%%%%%%%%%%%%%%%%%%%%%%%  

One naive candidate solution to the $\rho_c = 1/2$ task, which we will
contrast with the GA-evolved CAs, is the $r=3$ ``majority vote'' CA.
This CA, denoted \caT{maj}, maps the center cell in each $7$-cell
neighborhood to the majority state in that neighborhood.
\fig{majority} gives two space-time diagrams illustrating the behavior
of \caT{maj} on two ICs, one with $\rho_0 < 1/2$ and the other with
$\rho_0 > 1/2$.  As can be seen, small high-density (low-density)
regions are mapped to regions of all $1$s ($0$s). But when an all-$1$s
region and an all-$0$s region border each other, there is no way to
decide between them and both persist.  Thus, \caT{maj} does not
perform the $\rho_c = 1/2$ task.  In particular,
\performance{10^4}{N}{\caT{maj}} was measured to be zero for $N \in
\{149, 599, 999\}$. At a minimum more sophisticated coordination in
the form of information transfer and decision making is required. And,
given the local nature of control and communication in CAs, the
coordination among cells must emerge in the absence of any central
processor or central memory directing the cells.

Other researchers, building on our work, have examined variations of
the $\rho_c=1/2$ task that can be performed by simple CAs or by
combinations of CAs.  Capcarrere et al.\,\cite{CapcarrereEtAl96} noted
that changing the output specification makes the task significantly
easier. For example, ECA 184 classifies densities of initial conditions
within $\lceil N/2 \rceil$ time steps by producing a final configuration
of a checkerboard pattern $(01)^*$ interrupted by one or more blocks of
at least two consecutive $0$s for low-density ICs or at least two
consecutive $1$s for high-density ICs. Fuk\'{s}
\cite{Fuks97} noted that by using the final configuration of ECA 184
as the initial configuration of ECA 232, the correct final
configuration of either all-$0$s or all-$1$s is obtained. Note that
Fuk\'{s}' solution requires a central controller that counts time up
to $\lceil N/2 \rceil$ steps in order to shift from a CA using rule
184 to one using rule 232.

Both solutions always yield correct density classification, whereas the
single-CA $\rho_c=1/2$ task is considerably more difficult.  In fact,
it has been proven that no single, finite-radius two-state CA can
perform the $\rho_c=1/2$ task perfectly for all $N$ \cite{Das98,Land&Belew95}.

Our interest is not focused on developing a new and better parallel
method for
performing this specific task. Clearly, one-dimensional, binary-state
cellular automata are far from the best architectures to use if one is
interested in performing density classification efficiently. As we
have emphasized before, the task is trivial within other computational
model classes. Instead, our interest is in investigating how GAs can
design CAs that have interesting collective computational capabilities
and how we can understand those capabilities. Due to our more general
interest, we have been able to adapt this paradigm to other spatial
computation tasks---tasks for which the above specific solutions do
not apply and for which even approximate hand-designed CA solutions
were not previously known \cite{DasEtAl95a}.

%%%%%%%%%%%%%%%%%%%% EVOLVING CAs WITH GAs %%%%%%%%%%%%%%%%%%%%%%%%%%%

\section{Evolving Cellular Automata with Genetic Algorithms 
\label{evolving-ca}}

Genetic algorithms are search methods inspired by biological evolution
\cite{Holland92a}. In a typical GA, candidate solutions to a given
problem are encoded as bit strings.  A population of such strings
(``chromosomes'') is chosen at random and evolves over several
generations under selection, crossover, and mutation.  At each
generation, the fitness of each chromosome is calculated according to
some externally imposed fitness function and the highest-fitness
chromosomes are selected preferentially to form a new population via
reproduction.  Pairs of such chromosomes produce offspring via
crossover, where an offspring receives components of its chromosome
from each parent. The offspring chromosomes are then subject at each
bit position to a small probability of mutation (i.e., being flipped). 
After several generations, the population often contains high-fitness
chromosomes---approximate solutions to the given problem.  (For
overviews of GAs, see \cite{Goldberg89c,Mitchell96a}.)

We used a GA to search for $(k,r) = (2,3)$ CAs to perform the $\rho_c = 1/2$
task.\footnote{The basic framework was introduced in Ref. \cite{Packard89}
to study issues of phase transitions, computation, and adaptation. For a review
of the original motivations and a critique of the results see 
Ref. \cite{MitchellEtAl93a}.} Each chromosome in the population
represented a candidate CA---it consisted of the output bits of
the rule table, listed in lexicographic order of neighborhood (cf.
$\phi$ in \fig{ca-picture}). The chromosomes representing CAs
were of length $2^{2r+1}=128$ bits. The size of the 
space in which the GA searched was thus $2^{128}$---far too large for
exhaustive enumeration and performance evaluation.

Our version of the GA worked as follows.  

First, an initial population of $M$ chromosomes was chosen at random.
The {\it fitness} \fitness{I}{N}{$\phi$} of a CA $\phi$ in the
population was computed by randomly choosing $I$ ICs on a lattice of
$N$ cells, iterating the CA on each IC either until it arrived at a fixed
point or for a maximum of $T_{\textsub{max}}$ time steps. It was then
determined whether the final configuration was correct---i.e., the
all-$0$s fixed point for $\rho_0 < 1/2$ or the all-$1$s fixed point
for $\rho_0 > 1/2$. \fitness{I}{N}{$\phi$} was the fraction of the
$I$ ICs on which $\phi$ produced the correct final behavior. No
credit was given for partially correct final configurations.

In each generation, (1) a new set of $I$ ICs was generated; (2)
\fitness{I}{N}{$\phi$} was computed for each CA $\phi$ in the
population; (3) CAs in the population were ranked in order of fitness
(with ties broken at random); (4) a number $E$ of the highest fitness
CAs (the ``elite'') was copied to the next generation without
modification; and (5) the remaining $M-E$ CAs for the next generation
were formed by crossovers between randomly chosen pairs of the elite
CAs. With probability $p_c$, each pair was crossed over at a single
randomly chosen locus $l$, forming two offspring.  The first child
inherited bits $0$ through $l$ from the first parent and bits $l+1$
through $127$ from the second parent; vice versa for the second child.
The parent CAs were chosen for crossover from the elite with
replacement---that is, an elite CA was permitted to be chosen any
number of times. The two offspring chromosomes from each crossover (or
copies of the parents, if crossover did not take place) were mutated
($0 \rightarrow 1$ and $1 \rightarrow 0$) at each locus with
probability $p_m$. This process was repeated for $G$ generations
during a single GA run. Note that since a different sample of ICs was
chosen at each generation, the fitness function itself is a random
variable.

We ran experiments with two different distributions for choosing the
$M$ chromosomes in the initial population and the set of $I$ ICs at
each generation: (i) an ``unbiased'' distribution in which each bit's
value is chosen independently with equal probability for 0 and 1, and
(ii) a density-uniform distribution in which strings were chosen with
uniform probability over $\lambda \in [0,1]$ or over $\rho_0 \in
[0,1]$, where $\lambda$ is the fraction of $1$s in $\phi$'s output
bits and $\rho_0$ is the fraction of $1$s in the IC. Using the
density-uniform distribution for the initial CA population and for the
ICs considerably improved the GA's ability to find high fitness CAs on
any given run. (That is, we could use $50\%$ fewer generations per GA run
and still find high performance CAs.) The results we report here are
from experiments in which density-uniform distributions were used.

The experimental parameters we used were $M=100$, $I=100$, $E=20$, $N
= 149$, $T_{\textsub{max}} = 2N$, $p_c = 1.0$ (i.e., crossover was
always performed), $p_m = 0.016$, and $G=100$.  Experiments using
variations on these parameters did not result in higher performance
solutions or faster convergence to the best-performance solutions.

To test the quality of the evolved CAs we used \perf{10^4}{N} with
$N \in \{149, 599, 999\}$. This performance measure is a more stringent
quality test than the fitness \fit{100}{N} used in the GA runs: under \perf{10^4}{N}
the ICs are chosen from an unbiased distribution and thus have $\rho_0$
close to the density threshold $\rho = 1/2$. Such ICs are the hardest cases
to classify. Thus, \perf{10^4}{N} gives a lower bound on other performance
measures.  In machine learning terms, the ICs used to calculate \fit{100}{149}
are the training sets for the CAs and the ICs used to calculate \perf{10^4}{N}
are larger and harder test sets that probe the evolved CA's
generalization ability.  

%%%%%%%%%%%%%%%%%%%% RESULTS OF EXPERIMENTS %%%%%%%%%%%%%%%%%%%%%%%%%%%

\section{Results of Experiments \label{results-of-experiments}}

In this section we describe the results from 300 independent runs of
this GA, with different random number seeds.

In each of the 300 runs, the population converged to CAs implementing
one of three types of computational strategies.  The term ``strategy''
here refers to the mechanisms by which the CA attains some level of
fitness on the $\rho_c=1/2$ task. These three strategy types,
``default'', ``block expanding'', and ``particle'', are illustrated
in \figs{defAB}--\ref{parAB}. In each figure, each row contains two
space-time diagrams displaying the typical behavior of a CA $\phi$
that was evolved in a GA run. Thus, CAs from six different runs are
shown. In each row, $\rho_0 < 1/2$ in the left space-time diagram and
$\rho_0 > 1/2$ in the right. The rule tables and measured \perf{10^4}{N}
values for the six CAs are given in Table~\ref{ca-table}.

\subsection{Default Strategies}

\begin{figure}
\centerline{\psfig{figure=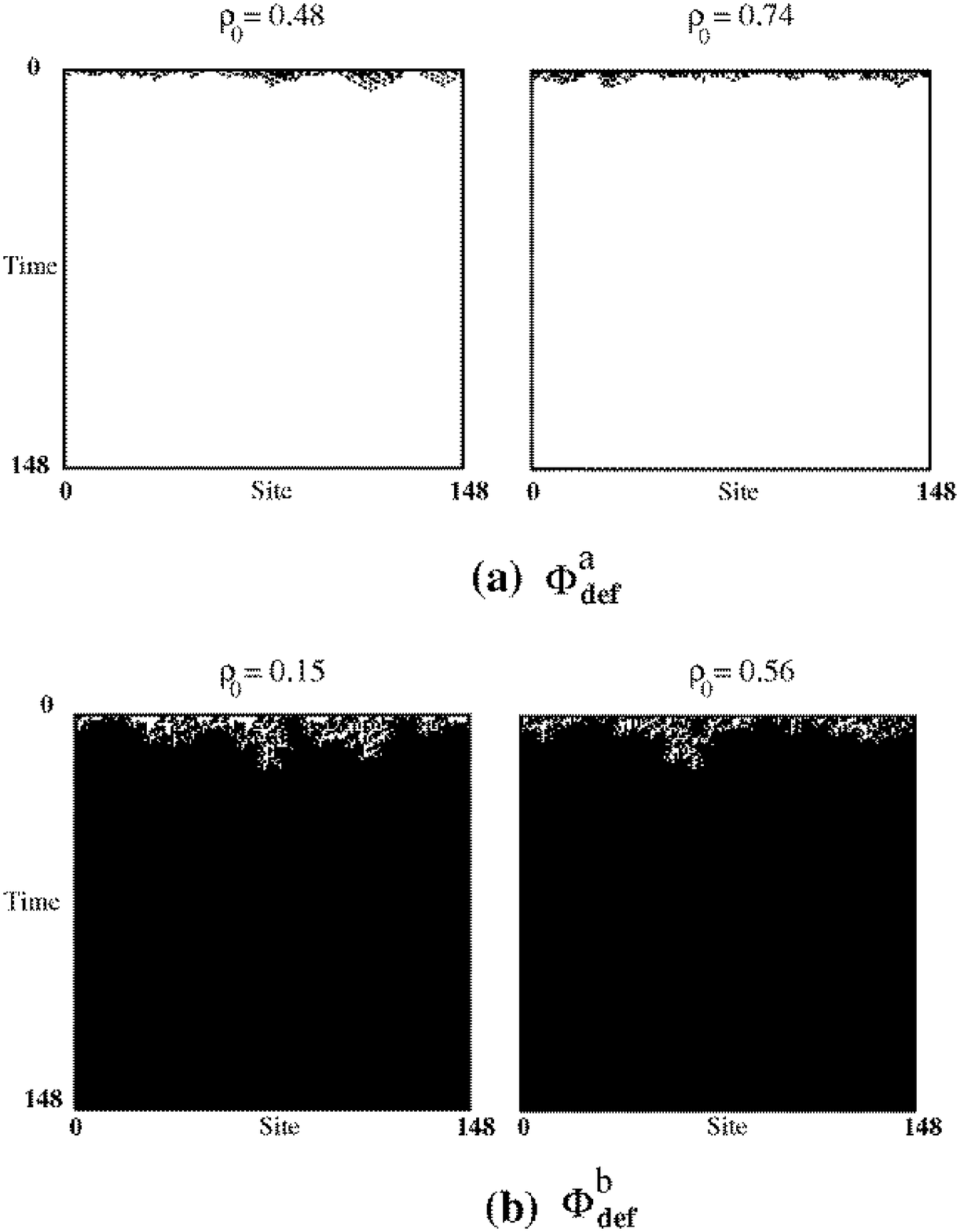,height=7in}}
\caption{Space-time behavior of ``default'' strategy CAs evolved on
  two different GA runs. (a) \caS{def}{a} with $\rho_0 = 0.48$ (left)
  and $\rho_0 = 0.74$ (right). On almost all ICs this CA iterates to a
  fixed point of all $0$s, correctly classifying only low-$\rho$ ICs.
  (b) \caS{def}{b} with $\rho_0 = 0.15$ (left) and $\rho_0 = 0.56$
  (right). On almost all ICs this CA iterates to a fixed point of all
  $1$s, correctly classifying only high-$\rho$ ICs. 
\label{defAB}}
\end{figure}

In 11 out of the 300 runs, the highest performance CAs implemented
``default'' strategies, which on almost all ICs iterate to all $0$s or
all $1$s, respectively.  The typical behavior of two such CAs,
\caS{def}{a} and \caS{def}{b}, is illustrated in \figs{defAB}(a) and
\ref{defAB}(b).  Default strategies each have \performance{I}{N}{$\phi$} $\approx
0.5$, since each classifies one density range (e.g., $\rho < 1/2$)
correctly and the other ($\rho > 1/2$) incorrectly. Since the initial
CA population is generated with uniform distribution over $\lambda$
it always contains some CAs with very high or low $\lambda$. And since
$\lambda$ is the fraction of $1$s in the output bits of the look-up
table, these extreme-$\lambda$ CAs tend to have one or the other
default behavior.

%%%%%%%%%%%%%%%%%%% DEF %%%%%%%%%%%%%%%%%%%  

%%%%%%%%%%%%%%%%%%%%%%%%%%%%%%%%%%%%%%%%%%%%%%%%%%%%%%%%%%%%%%

\subsection{Block-Expanding Strategies}

\begin{figure}
\centerline{\psfig{figure=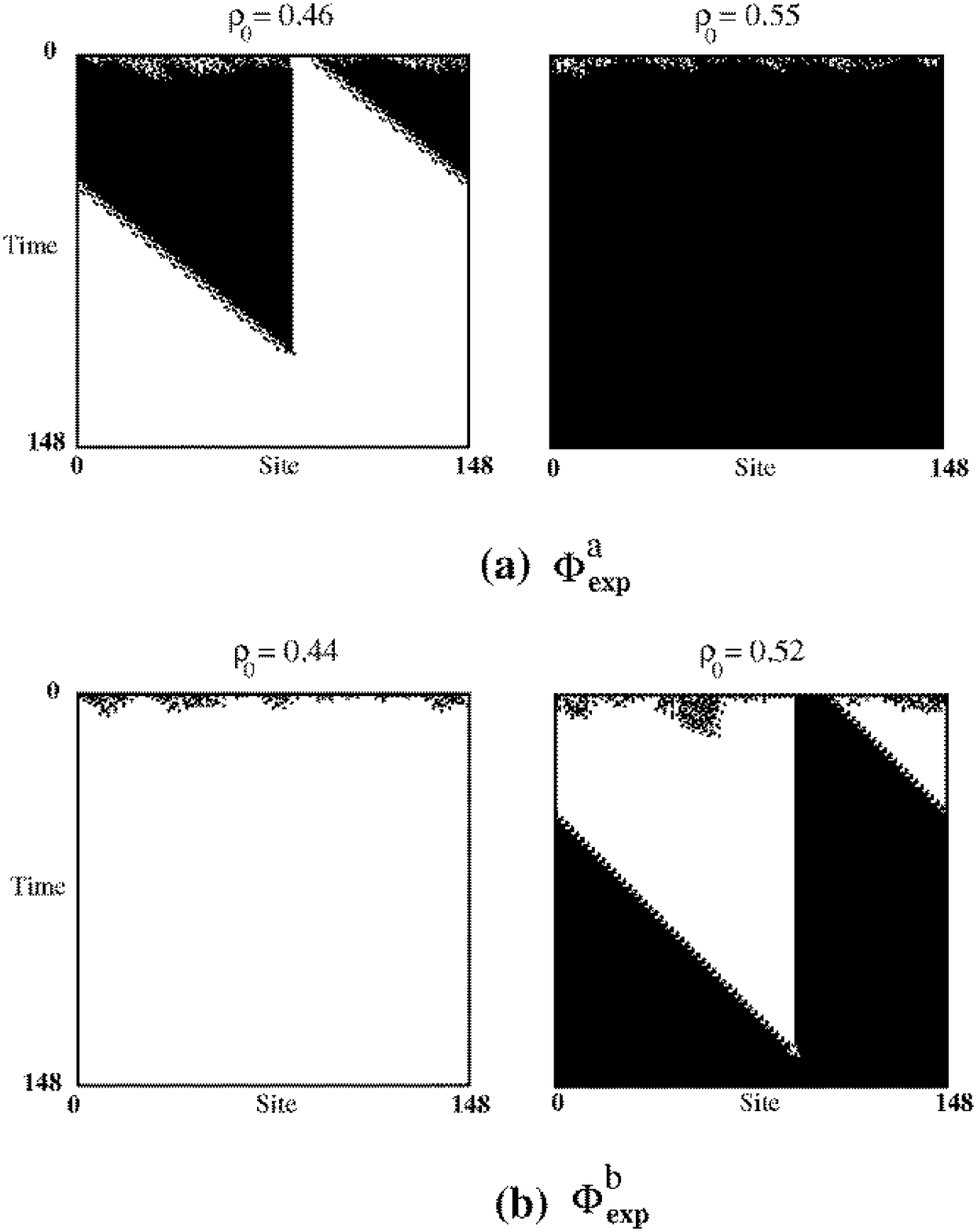,height=7in}}
\caption{Space-time behavior of two ``block expanding'' CAs evolved on 
  different GA runs. (a) \caS{exp}{a} with $\rho_0 = 0.46$ (left) and
  $\rho_0 = 0.55$ (right).  This CA defaults to a fixed point of all
  $1$s unless the IC contains a sufficiently large block of adjacent
  $0$s, in which case, that block is expanded. (b) \caS{exp}{b} with
  $\rho_0 = 0.44$ (left) and $\rho_0 = 0.52$ (right). This CA defaults
  to a fixed point of all $0$s unless the IC contains a sufficiently
  large block of adjacent $1$s, in which case, that block is expanded.
  The classification of the IC is correct in each of these four cases.
\label{expAB}}
\end{figure}

In most runs (280 out of 300 in our experiments) the GA evolved CAs
with strategies like those shown in \figs{expAB}(a) and \ref{expAB}(b).
\caS{exp}{a} in \fig{expAB}(a) defaults to an all-$1$s fixed point
(right diagram) unless there is a sufficiently large block of adjacent
(or almost adjacent) $0$s in the IC. In this case it expands that
block until $0$s fill up the entire lattice (left diagram).
\caS{exp}{b} in \fig{expAB}(b) has the opposite strategy. It defaults
to the all-$0$s fixed point unless there is a sufficiently large block
of $1$s in the IC. The meaning of ``sufficiently large block'' depends
on the particular CA, but is typically close to the neighborhood size
$2r+1$. For example, \caS{exp}{a} will expand blocks of 8 or more $0$s
and \caS{exp}{b} will expand blocks of $7$ or more $1$s.

These ``block-expanding'' strategies rely on the presence or absence
of blocks of $1$s or $0$s in the IC: blocks of adjacent $0$s ($1$s)
are more likely to appear in low- (high-) density ICs.  Since the
occurrence of such blocks is statistically correlated with $\rho_0$,
recognizing and then expanding them leads to fitnesses above those for
the default strategy. The strength of this correlation depends on the
initial density $\rho_0$ and on the lattice size $N$.  Typical
block-expanding strategies have \fit{100}{149} $\approx 0.9$ and
\perf{10^4}{149} $\approx 0.6$. The block-expanding strategies designed
by the GA are adapted to $N=149$; their performances do not scale well
to larger lattice sizes. This occurs since the probability of a block
of, say, seven adjacent $1$s appearing for a given $\rho_0$ increases
with $N$ and this means that the correlation between the occurrence of
this block and density decreases. This can be seen in the measured
values of \perf{10^4}{N} for \caS{exp}{a} and \caS{exp}{b} for longer
lattices given in Table~\ref{ca-table}.

%%%%%%%%%%%%%%%%%%% EXP %%%%%%%%%%%%%%%%%%%  

%%%%%%%%%%%%%%%%%%%%%%%%%%%%%%%%%%%%%%%%%%%%%%%%%%%%%%%%%%%%%%

\subsection{Embedded-Particle Strategies}

\begin{figure}
\centerline{\psfig{figure=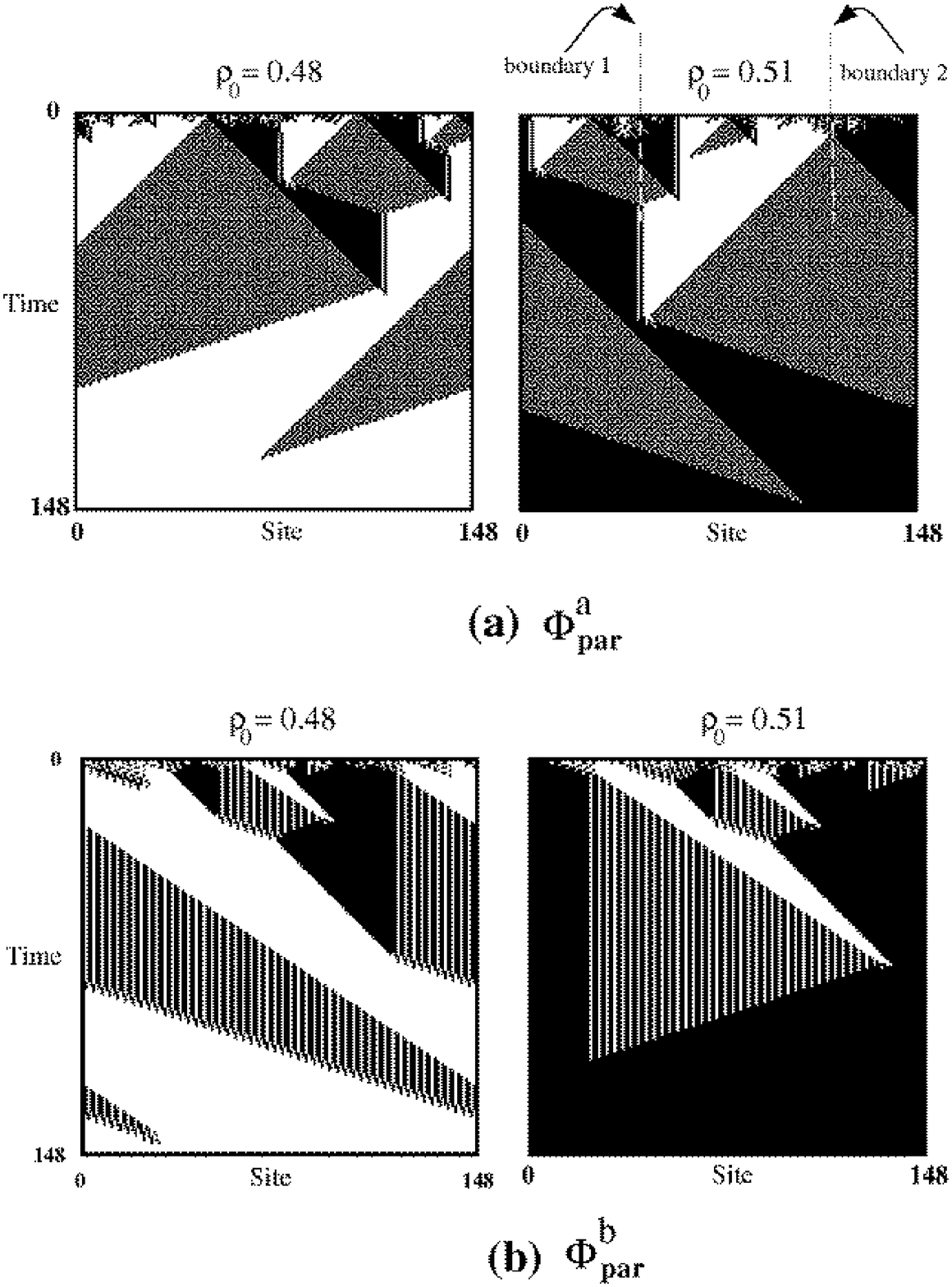,height=7in}}
\caption{
  Space-time behavior of two ``particle'' CAs evolved on different GA
  runs.  (a) \caS{par}{a} with $\rho_0 = 0.48$ (left) and $\rho_0 =
  0.51$ (right).  (b) \caS{par}{b} with $\rho_0 = 0.48$ (left) and
  $\rho_0 = 0.51$ (right). These CAs use the boundaries between
  homogeneous space-time regions to effect information transmission
  and processing.  Again, the classification of the IC is correct in
  each of these four cases.
  \label{parAB}}
\end{figure}

The block-expanding strategies are not examples of the kind of sophisticated
coordination and information transfer that we claimed must be achieved for
robust performance on the $\rho_c=1/2$ task. Under these strategies all
the computation is done locally in identifying and then expanding a
``sufficiently large'' block. Moreover, the performance on $N = 149$
does not generalize to larger lattices. Clearly, the block-expanding
strategies are missing important aspects required by the task. The third class
of strategies evolved by the GA, the ``embedded-particle'' strategies,
do achieve the coordination and communication we alluded to earlier. Typical
space-time behaviors of two particle strategies, \caS{par}{a} and \caS{par}{b},
are given in \figs{parAB}(a) and \ref{parAB}(b). It can be seen that there is a transient
phase during which the spatial and temporal transfer of information about the
local density takes place. Such strategies were evolved in $9$ out of the $300$
runs. 

\caS{par}{a}'s behavior is somewhat similar to that of \caT{maj} in that local
high-density regions are mapped to all $1$s and local low-density regions
are mapped to all $0$s. In addition, a vertical stationary boundary
separates these regions. The set of local spatial configurations that
make up this boundary is specified in formal language terms by the
regular expression $(1)^+01(0)^+$, where $(w)^+$ means a
positive number of repetitions of the word $w$ \cite{Hopc79}.

The stationary boundary appears when a region of $1$s on the left meets
a region of $0$s on the right. However, there is a crucial difference from
\caT{maj}: when a region of $0$s on the left meets a region of $1$s on
the right, a checkerboard region $(01)^+$ grows in size with equal
speed in both directions. A closer analysis of its role in the overall
space-time behavior shows that the checkerboard region serves to decide
which of the two adjacent regions ($0$s and $1$s) is the larger. It does
this by simply cutting off the smaller region and so the larger ($0$
or $1$) region continues to expand. The net decision is that the density
in the region was in fact below or above $\rho_c = 1/2$. The spatial
computation here is largely geometric: there is a competition between the
sizes of high- and low-density regions.

For example, consider the right-hand space-time diagram of
\fig{parAB}(a).  The large low-density region between the lines marked
``boundary 1'' and ``boundary 2'' is smaller than the large high-density
region between ``boundary 2'' and ``boundary 1'' (moving left from
boundary 2 and wrapping around).  The left-hand side of the checkerboard
region (centered around boundary 2) collides with boundary 1 before the
right-hand side does. The result is that the collision cuts off the
inner white region, letting the outer black region propagate. 

In this way, \caS{par}{a}  uses local interactions and simple geometry to
determine the relative sizes of adjacent low- and high-density regions
that are larger than the neighborhood size. As is evident in \figs{parAB}(a)
and \ref{parAB}(b), this type of size competition over time happens across
increasingly larger spatial scales, gradually resolving competitions
between larger and larger regions.

The black-white boundary and the checkerboard region can be thought of as
signals indicating ``ambiguous'' density regions. Each of these boundaries
has local density exactly at $\rho_c = 1/2$. Thus, they are not themselves
``classified'' by the CA as low or high density. The result is that these
signals can persist over time.  The creation and interaction of these
signals can be interpreted as the locus of the computation being
performed by the CA---they form its emergent ``algorithm'', what we
have been referring to as the CA's ``strategy''.

\caS{par}{b} (\fig{parAB}(b)) follows a similar strategy, but with a vertically
striped region playing the role of the checkerboard region in \caS{par}{a}.
However, in this case there are asymmetries in the speeds of the
propagating region boundaries.  This difference yields a
lower \perf{10^4}{N}, as can be seen in Table~\ref{ca-table}.

These descriptions of the computational strategies evolved by the GA
are informal.  A major goal of our work is to make terms such as
``computation'', ``computational strategy'', and ``emergent
algorithm'' more rigorous for cellular automata.  In the next section
we will describe how we are using the notions of domains, particles,
and particle interactions to do this. We will use these notions to
answer questions such as, How, precisely, is a given CA performing
the task? What structural components are used to support this
information processing? How can we predict \perf{I}{N} and other
computational properties of a given CA? Why is \perf{I}{N} greater
for one CA than for another? What types of mistakes does a given CA
make in performing the $\rho_c = 1/2$ task? These types of questions are
difficult, if not impossible, to answer in terms of local space-time
notions such as the bits in a CA's look-up table or even the raw
space-time configurations produced by the CA.  A higher-level
description is needed, one that incorporates computational structures.

%%%%%%%%%%%%%%%%%%% PAR %%%%%%%%%%%%%%%%%%%  

% CAs used in this figure

%parA id 64.72 in run 100
%00000101000001000000010110000110000001010000000000001111011101110000001101110111010101011000011101111011111111111011011101111111 1.000000

%parB id 99.57 in run 112580
%00000000000100100000000000110011010100000101000000010001001000110011101101110111111101111111111111111101111111111101010101111111

% ICs used for this figure: 
 
%parA_low
%01010100101101010010001011110100000000100101001100111010010110111001010110000111010100000011011011001000010111010111111001111111001000011000000111011
%
%parA_high
%10110000000100100010101011010101111100000110111011001111110100100000000010101101111001110000011001100111000101110010110001100011111100111010101111011
%
%parB_low
%00010100101000101001000100001101011011001111101010101010100100100000101101100100111110010011001111110111101100111010010100010101000111001100000001110
%
%parB_high
%11111110111000010001100100000010011001010010100110001011010100111101111000100100010010011100011011101011110001010110111010011111001011100010111001100

%%%%%%%%%%%%%%%%%%%%%%%%%%%%%%%%%%%%%%%%%%%%%%%%%%%%%%%%%%%%%%

%%%%%%%%%%%%%%%%%%% CA-TABLE %%%%%%%%%%%%%%%%%%%  

\begin{table}
\begin{center}
\begin{tabular}{|c|c|c|c|c|}
\hline
CA Name  &  Rule Table (Hexadecimal) &  \perf{10^4}{149} & \perf{10^4}{599} & \perf{10^4}{999} \\ \cline{1-5}
\caS{def}{a}   & {\tt 100111215030114D} & 0.500 & 0.500 & 0.500 \\
               & {\tt 01613507143B05BF} &       &       &       \\ \cline{1-5}
\caS{def}{b}   & {\tt 0BF9D97AF26F4F4B} & 0.499 & 0.499 & 0.501 \\
               & {\tt F3FF301F0B110DF7} &       &       &       \\ \cline{1-5}
\caS{exp}{a}   & {\tt 1010614604273F9B} & 0.656 & 0.523 & 0.504 \\ 
               & {\tt 7FD7D9DF35F53FFF} &       &       &       \\ \cline{1-5}
\caS{exp}{b}   & {\tt 02330A4B07016711} & 0.643 & 0.513 & 0.502 \\ 
               & {\tt 42D080C3CD877B7F} &       &       &       \\ \cline{1-5}
\caS{par}{a}   & {\tt 0504058605000F77} & 0.775 & 0.740 & 0.728 \\ 
               & {\tt 037755877BFFB77F} &       &       &       \\ \cline{1-5}
\caS{par}{b}   & {\tt 00240066A0A02246} & 0.766 & 0.687 & 0.641 \\ 
               & {\tt 76EFEFFFFBFFAAFE} &       &       &       \\ \cline{1-5}
\end{tabular} 
\end{center}
\caption{CA chromosomes (look-up table output bits) given in
hexadecimal and \perf{10^4}{N} for the six CAs illustrated in
\figs{defAB}--\ref{parAB}, on lattices of sizes $N=149$, $N=599$, and $N=999$.
To recover the 128-bit string giving the CA look-up table outputs, expand
each hexadecimal digit (left to right, top row followed by bottom row)
to binary.  This yields the neighborhood outputs in lexicographic
order of neighborhood, with the leftmost bit of the 128-bit string
giving the output bit for neighborhood {\tt 00000000}, and so on.
Since \perf{10^4}{N} is measured on a randomly chosen sample of ICs, it is a
random variable. This table gives its mean over 100 trials for each CA. 
Its standard deviation over the same 100 trials is approximately 0.005 for
each CA for all three values of $N$. For comparison, the best known
$(k,r) = (2,3)$ CAs for the $\rho_c = 1/2$ task have \perf{10^4}{149}
$\approx 0.85$ (see Sec.~\ref{related-work}). This appears to be close to
the upper limit of \perf{10^4}{149} for this class of spatial architectures.
\label{ca-table}}
\end{table}

%%%%%%%%%%%%%%%%%%%%%%%%%%%%%%%%%%%%%%%%%%%%%%%%%%%% 

%%%%%%% UNDERSTANDING COLLECTIVE COMPUTATION IN CELLULAR AUTOMATA %%%%%%%%%%%

\section{Understanding Collective Computation in Cellular Automata}

In this section we will describe our approach to formalizing the notion
of computational strategy in cellular automata and in other spatially
extended systems.  This approach is based on the computational
mechanics framework of Crutchfield \cite{Crut92c}, as applied to
cellular automata by Crutchfield and Hanson
\cite{Crut93a,Hanson93,Hans90a}. This framework comprises
a set of methods for classifying the different patterns that appear
in CA space-time behavior, using concepts from computation and dynamical
systems theories. These methods were developed as a way of analyzing the
behavior of cellular automata and other dynamical systems. They extend
more traditional geometric and statistical analyses by revealing the
intrinsic information-processing structures embedded in dynamical processes.

%%%%%%%% COMPUTATIONAL MECHANICS OF CELLULAR AUTOMATA %%%%%%%%%%%%

\subsection{Computational Mechanics of Cellular Automata}

As applied to cellular automata, the purpose of computational mechanics
is to discover an appropriate ``pattern basis'' with which to describe
the structural components that emerge in a CA's space-time behavior.
A CA pattern basis consists of a set ${\bf \Lambda}$ of formal languages
$\{\Lambda^i, i = 0,1,\ldots\}$ in terms of which a CA's space-time behavior
can be decomposed concisely and in a way constrained by the temporal dynamics. 
Once such a pattern basis is found, those cells in space-time regions
that are described by the basis can be seen as forming background ``domains''
against which coherent structures---defects, walls, etc.---not
fitting the basis move. In this way, structural features above and
beyond the domains can be identified and their dynamics analyzed
and interpreted on their own terms.

%%%%%%%%%%%%%%%%%%% ECA18 %%%%%%%%%%%%%%%%%%%  

\begin{figure}
\centerline{\psfig{figure=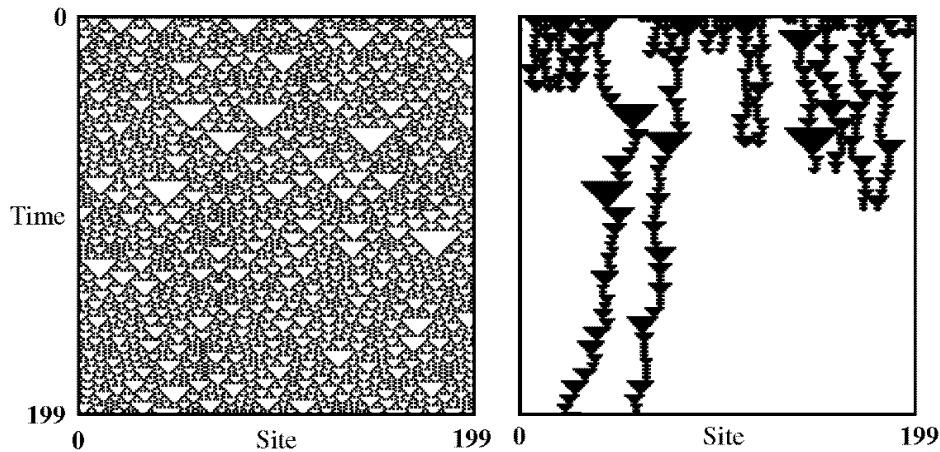,height=3in}}
\caption{(a) Space-time diagram illustrating the typical behavior of
  ECA 18---a CA exhibiting apparently random behavior, i.e., the set
  of length-$L$ spatial words has a positive entropy density as $L
  \rightarrow \infty$. (b) The same diagram with the regular
  domains---instances of words in \domain{0}---filtered out, leaving
  only the embedded particles ${\bf P} = \{1(00)^n1, n=0,1,2,\ldots
  \}$.  (After Ref. \cite{Crut92a}.)
\label{eca18}}
\end{figure}

%%%%%%%%%%%%%%%%%%%%%%%%%%%%%%%%%%%%%%%%%%%%%%%%%%%  

For example, consider the space-time diagram of \fig{eca18}(a),
illustrating the apparently random behavior of ECA 18. This example is
a useful illustration of embedded information processing since the
coherent structures are not immediately apparent to the eye. The
computational mechanics analysis \cite{Crut92a,Hans90a} of ECA 18 uses
a pattern basis consisting of the single domain language ${\bf
  \Lambda} = \{ \Lambda^0 = (0\Sigma)^+ \}$, where $\Sigma = \{ 0 , 1
\}$. That is, over most regions in ECA 18's configurations, every
other site is a $0$ and the remaining sites are wildcards, either $0$
or $1$. (Often this type of formal-language description of a set of
configuration features can be discovered automatically via the
``$\epsilon$-machine reconstruction'' algorithm
\cite{Crut92c,Hanson93}.)

Crutchfield and Hanson define a {\it regular domain} \domain{j} as a
space-time region that (i) is a regular language and (ii) is space- and
time-translation invariant. Regular domains can be represented by either
the set $\Lambda^i$ of configurations or by the minimal finite-state
machine ${\rm M}(\Lambda^i)$ that recognizes $\Lambda^i$. More specifically,
let $\{\Lambda^i \}$ be the pattern basis for CA $\phi$. Then the regular
domain \domain{i} describes the space-time regions of
$\{ \Phi^t ({\bf s}^0) : t = 0,1,2,\ldots\}$
whose configurations are words in \domain{i}. Formally, then, a regular
domain \domain{i} is a set that is
\begin{enumerate}
\item temporally invariant---the CA always maps a configuration in
\domain{i} to another configuration in \domain{i}:
$\Phi ({\bf s}) = {\bf s}^\prime,~{\bf s},{\bf s}^\prime \in \Lambda^i$; and
\item spatially homogeneous---the same pattern can occur at any site:
the recurrent states in the minimal finite automaton
${\rm M}(\Lambda^i)$ recognizing $\Lambda^i$ are strongly connected.
\end{enumerate}

Once a CA's regular domains are discovered, either through visual inspection
or by an automated induction method, and proved to satisfy the above two
conditions, then the corresponding space-time regions are, in a sense,
understood.  Given this level of discovered regularity, the domains can be
filtered out of the space-time diagram, leaving only the ``unmodeled''
deviations, referred to as domain ``walls'', whose dynamics can then be
studied in and of themselves. Sometimes, as is the case
for the evolved CA we analyze here, these domain walls are spatially
localized, time-invariant structures and so can be considered to be
``particles''.

In ECA 18 there is only one regular domain \domain{0}. It turns out
that it is stable and so is called a regular ``attractor''---the
stable invariant set to which configurations tend over long times,
after being perturbed away from it by, for example, flipping a site
value \cite{Crut92a,Hans90a}.
Although there are random sites in the domain, its basic pattern
is described by a simple rule: all configurations are allowed in
which every other site value is a $0$. If these fixed-value sites are
on even-numbered lattice sites, then the odd-numbered lattice sites
have a wild card value, being $0$ or $1$ with equal probability.
The boundaries between these ``phase-locked'' regions are ``defects''
in the spatial periodicity of \domain{0} and, since they are spatially
localized in ECA 18, they can be thought of as particles embedded in the
raw configurations.

To locate a particle in a configuration generated by ECA 18, assuming
one starts in a domain, one scans across the configuration, from left to
right say, until the spatial period-$2$ phase is broken. This occurs
when a site value of $1$ is seen where the domain pattern indicates a
$0$ should be. Depending on a particle's structure, it can occur, as it
does with ECA 18, that scanning the same configuration in the opposite
direction (right to left) may lead to the detection of the broken domain
pattern at a different site. In this case the particle is defined to be
the set of local configurations between these locations.

Due to this ECA 18's particles are manifest in spatial configurations as
blocks in the set ${\bf P} = \{ 1(00)^n1, n = 0, 1, 2, \ldots\}$, a
definition that is left-right scan invariant. \fig{eca18}(b) shows a
filtered version of \fig{eca18}(a) in which the cells participating in
\domain{0} are colored white and the cells participating in $\bf P$ are
colored black. The spatial structure of the particles is reflected in
the triangular structures, which are regions
of the lattice in which the particle---the breaking of \domain{0}'s
pattern---is localized, though not restricted to a single site.

%%%%%%%%%%%%%%%%% ECA 18 Particle Catalog %%%%%%%%%%%%%%%%%%%%%%

\begin{table}
\begin{center}
\begin{tabular}{|c|}
\hline 
{\bf Regular Domain}                    \\ \hline
\domain{0} $=\{ (0(0+1))^* \}$           \\ \hline
{\bf Particle}                          \\ \hline
$\alpha \sim$ \domain{0} \domain{0} $=\{ 1(00)^n1, n = 0,1,2, \ldots \}$    \\ \hline
{\bf Interaction} (annihilation)        \\ \hline
$\alpha + \alpha \rightarrow \emptyset ~~ (\Lambda^0)$ \\ \hline
\end{tabular}
\end{center}
\caption{ECA 18's catalog of regular domains, particles, and particle
interactions. The notation $p \sim \Lambda^i\Lambda^j$ means that $p$
is the particle forming the boundary between domains \domain{i} and
\domain{j}.
\label{ECA18ParticleCatalogl}} 
\end{table}
%%%%%%%%%%%%%% END of PARTICLE TABLE A %%%%%%%%%%%%%%%%%%%

In this way, ECA 18's configurations can be decomposed into natural,
``intrinsic'' structures that ECA 18 itself generates; viz., its domain
\domain{0} and its particle ${\bf P}$. These structures are summarized
for a CA in what we call a {\em particle catalog}. The catalog is
particularly simple for ECA 18; cf. Table \ref{ECA18ParticleCatalogl}.
The net result is that ECA 18's behavior can be redescribed at the
higher level of particles. It is noteworthy that, starting from
arbitrary initial configurations, ECA 18's particles have been shown to
follow a random walk in space-time on an infinite lattice, annihilating
in pairs whenever they intersect \cite{Eloranta94,Hans90a}. One
consequence is that there is no further structure, such as coherent
particle groupings, to understand in ECA 18's dynamics. Thus, one moves
from the deterministic dynamics at the level of the CA acting on raw
configurations to a level of stochastic particle dynamics. The result
is that ECA 18 configurations, such as those in \fig{eca18}(a), can be
analyzed in a much more structural way than by simply classifying ECA
18 as ``chaotic''.

In the computational mechanics view of CA dynamics, embedded particles
carry various kinds of information about local regions in the IC.
Given this, particle interactions are the loci at which this
information is combined and processed and at which decisions are made.
In general, these structural aspects---domains, particles, and
interactions---do not appear immediately.  As will be seen below,
often there is a initial disordered period, after which the
configurations condense into well-defined regular domains, particles,
and interactions. To capture this relaxation process we define the
{\it condensation time} $t_{\textsub{c}}$ as the first iteration at
which the filtered space-time diagram contains only well-defined
domains in $\bf \Lambda$ and the walls between them. In other words,
at $t_{\textsub{c}}$, every cell participates in either a regular
domain, of width at least $2r+1$, in a wall between them, or in an
interaction between walls. (See Refs. \cite{CrutchfieldEtAl97a} and
\cite{HordijkEtAl98} for a more detailed discussion
of the condensation phase and its consequences.)

%%%%%%% COMPUTATIONAL MECHANICS OF EVOLVED CELLULAR AUTOMATA %%%%%%%

\subsection{Computational Mechanics of Evolved Cellular Automata}

This same methodology is particularly useful in understanding and
formalizing the computational strategies that emerged in the
GA-evolved CA. Fortunately, in the following exposition most of the
structural features in the evolved CA are apparent to the eye.
\fig{parAB}(a) suggests that an appropriate pattern basis for
\caS{par}{a} is ${\bf \Lambda} = \{\Lambda^0 = 00^+, \Lambda^1 = 11^+,
\Lambda^2 = (01)^+\}$, corresponding to the all-white, all-black, and
checkerboard regions.  Similarly, \fig{parAB}(b) suggests that for
\caS{par}{b} we use ${\bf \Lambda} = \{ \Lambda^0 = 00^+, \Lambda^1 =
111^+, \Lambda^0 = (011)^+\}$, corresponding to the all-white,
all-black, and striped regions.

Note that a simple shortcut can be used to identify domains that are
spatially and temporally periodic. If the same ``pattern'' appears
repeated over a sufficiently large ($\gg$ $r$ cells by $r$ time steps)
space-time region, then it is a domain. It is also particularly easy
to prove such regions are regular domains. Exactly how the pattern is
expressed as a regular language or as a minimal finite-state machine
typically requires closer inspection.

%%%%%%%%%%%%%%%%%%% PAR_AB-HIGH-FIL %%%%%%%%%%%%%%%%%%%  
\begin{figure}
\centerline{\psfig{figure=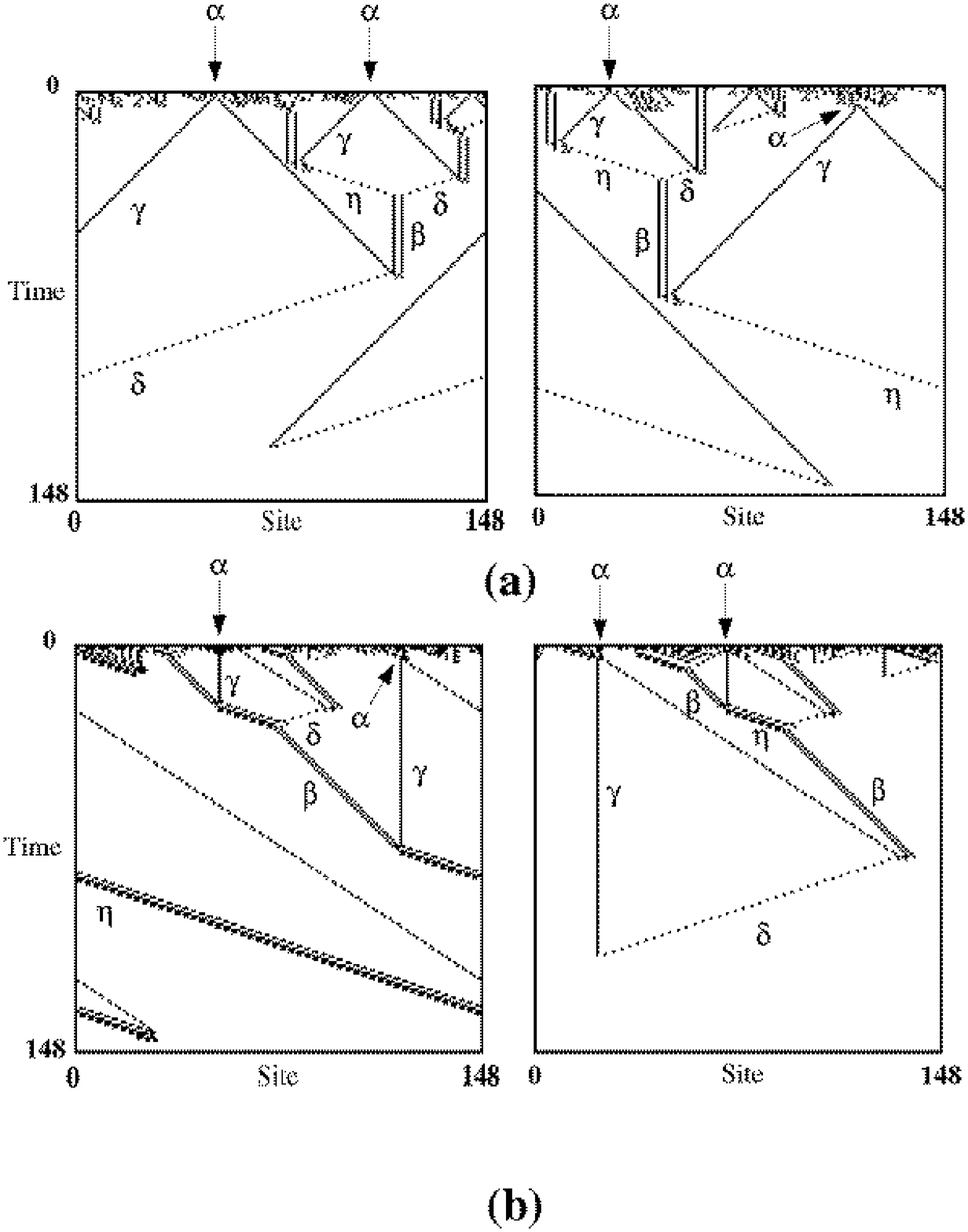,height=7in}}
\caption{(a) Version of \fig{parAB}(a) with the regular domains
filtered out, revealing the particles and their interactions.
(b) Filtered version of \fig{parAB}(b). 
\label{parAB-filtered}}
\end{figure}
%%%%%%%%%%%%%%%%%%%%%%%%%%%%%%%%%%%%%%%%%%%%%%%%%%%%%%%%

Once identified, the computational contributions of these space-time
regions can be easily understood. The contributions consist solely of
the generation of words in the corresponding regular language. Since
this requires only a finite amount of spatially localized memory, its
direct contribution to the global computation required by the task is
minimal. (The density of memory vanishes as the domain increases in
size.) The conclusion is that the domains themselves, while necessary,
are not the locus of the global information processing.

\fig{parAB-filtered} is a version of \fig{parAB} with \caS{par}{a}'s
and \caS{par}{b}'s
regular domains filtered out. The result reveals the walls between
them, which for \caS{par}{a} and \caS{par}{b} 
are several kinds of embedded particles.
The particles in \fig{parAB-filtered} are labeled with Greek letters.
This filtering is performed by a building a transducer that reads
in the raw configurations and can recognize when sites are in which
domain. The transducer used for \fig{parAB-filtered}(a), for example,
outputs white
at each site in one of \caS{par}{a}'s domains and black at each site
participating in a domain wall. (The particular transducer and comments
on its construction and properties can be found in Appendix
\ref{PhiParATransducer}. The general construction procedure is given
in Ref. \cite{Crut93a}.)

Having performed the filtering, the focus of analysis shifts away from the
raw configurations to the new level of embedded-particle structure. The
questions now become, Are the computational strategies explainable in terms of
particles and their interactions ? Or, is there as yet some unrevealed
information processing occurring that is responsible for high performance?

%%%%%%%%%%%%%%%%% PARTICLE TABLE A %%%%%%%%%%%%%%%%%%%%%%

\begin{table}
\begin{center}
\begin{tabular}{|c|c|c|}
\hline 
\multicolumn{3}{|c|}{{\bf Regular Domains}} \\ \cline{1-3}
\domain{0} $=\{0^+\}$ & \domain{1}$=\{1^+\}$ & \domain{2}$=\{(01)^+\}$ \\ \cline{1-3}
\multicolumn{3}{|c|}{{\bf Particles (Velocities)}} \\ \cline{1-3}
$\alpha \sim$ \domain{0} \domain{1} (---) &
$\beta \sim$ \domain{1} \domain{0} (0) &
% the "01" between \domain{1} & \domain{0} is a part of the particle, 
% and it need not be imposed a priori.  Other particles
% also have similar features. raja
$\gamma \sim$ \domain{0} \domain{2} (-1) \\
$\delta \sim$ \domain{2} \domain{0} (-3) &
$\eta \sim$ \domain{1} \domain{2} (3) & 
$\mu \sim$ \domain{2} \domain{1} (1) \\ \cline{1-3}
\multicolumn{3}{|c|}{{\bf Interactions}} \\ \cline{1-3}
decay   &   \multicolumn{2}{c|}{$\alpha \rightarrow \gamma + \mu$}  \\ 
react   &   \multicolumn{2}{c|}{$\beta + \gamma \rightarrow \eta$, $\mu + \beta \rightarrow \delta$, $\eta + \delta \rightarrow \beta$}  \\ 
annihilate &  \multicolumn{2}{c|} {$\eta + \mu \rightarrow \emptyset~~(\Lambda^1)$, $\gamma + \delta \rightarrow \emptyset~~(\Lambda^0)$} \\ \cline{1-3}
\hline
\end{tabular}
\end{center}
\caption{\caS{par}{a}'s catalog of regular domains, particles (including
velocities in parentheses), and particle interactions. Note that this
catalog leaves out possible three-particle interactions.
\label{particle-table_a}} 
\end{table}
%%%%%%%%%%%%%% END of PARTICLE TABLE A %%%%%%%%%%%%%%%%%%%

%%%%%%%%%%%%%% PARTICLE TABLE B %%%%%%%%%%%%%%%%%%%%%%%%%%

\begin{table}
\begin{center}
\begin{tabular}{|c|c|c|}
\hline 
\multicolumn{3}{|c|}{{\bf Regular Domains}} \\ \cline{1-3}
\domain{0} $=\{0^+\}$ & \domain{1} $=\{1^+\}$ & \domain{2} $= \{(011)^+\}$ \\ \cline{1-3}
\multicolumn{3}{|c|}{{\bf Particles (Velocities)}} \\ \cline{1-3}
$\alpha \sim$ \domain{1} \domain{0} (0) &
$\beta \sim$ \domain{0} \domain{1} (1) & 
$\gamma \sim$ \domain{1} \domain{2} (0) \\
$\delta \sim$ \domain{2} \domain{1} (-3) &
$\eta \sim$ \domain{0} \domain{2} (3) & 
$\mu \sim$ \domain{2} \domain{0} (3/2) \\ \cline{1-3}
\multicolumn{3}{|c|}{{\bf Interactions}} \\ \cline{1-3}
decay  &   \multicolumn{2}{c|}{$\alpha \rightarrow \gamma + \mu$}  \\ 
react  &   \multicolumn{2}{c|}{$\beta + \gamma \rightarrow \eta$, $\mu + \beta \rightarrow \delta$, $\eta + \delta \rightarrow \beta$}  \\ 
annihilate &      \multicolumn{2}{c|}{$\eta + \mu \rightarrow \emptyset~~(\Lambda^0)$, $\gamma + \delta \rightarrow \emptyset~~(\Lambda^1)$} \\ \cline{1-3}
\end{tabular}
\end{center}
\caption{\caS{par}{b}'s catalog of regular domains, particles (including
their velocities in parentheses), and particle interactions.
\label{particle-table_b}}
\end{table}
%%%%%%%%%%%%%% End of PARTICLE TABLE B %%%%%%%%%%%%%%%%%%%%%%

Tables~\ref{particle-table_a} and ~\ref{particle-table_b} list all the
different particles that are observed in the space-time behavior of
\caS{par}{a} and \caS{par}{b}, along with their velocities and the
interactions that can take place between them. Note that these particle
catalogs do not include all possible structures, for example, possible
three-particle interactions. The computational strategies of
\caS{par}{a} and \caS{par}{b} can now be analyzed in terms of the
particles and their interactions as listed in the particle catalogs.

%%%%%%%% COMPUTATIONAL STRATEGY OF PAR_A %%%%%%%%%%%%

\subsection{Computational Strategy of \caS{par}{a}}

In a high-performance CA such as \caS{par}{a}, particles carry information
about the density of local regions in the IC, and their interactions
combine and process this information, rendering a series of decisions
about $\rho_0$.  How do these presumed ``functional'' components lead
to the observed fitness and computation performance?

Referring to Table \ref{particle-table_a} and \fig{parAB-filtered}(a),
\caS{par}{a}'s $\beta$ particle is seen to consist of the zero-velocity
black-to-white boundary.  $\beta$ carries the information
that it came from a region in the IC in which the density is locally
ambiguous: the density of $1^k0^k$, when determined at its center,
is exactly $\rho_c$. The ambiguity cannot be resolved locally. It
might be, however, at a later time, when more information can be
integrated from other regions of the IC.

Likewise, the $\alpha$ ``particle'' consists of the white-to-black boundary,
but unlike the $\beta$ particle, $\alpha$ is unstable and immediately
decays into two particles $\gamma$ (white-checkerboard boundary) and
$\mu$ (checkerboard-black boundary). Like $\beta$, $\alpha$ indicates
local density ambiguity. The particles into which it decays, $\gamma$ and $\mu$,
carry this information and so they too are ``ambiguous density'' signals. 
$\gamma$ carries the information that it borders a white (low density) region
and $\mu$ carries the information that it borders a black (high density)
region. The two particles also carry the mutual information of having come
from the same ambiguous density region where the transient $\alpha$ was
originally located. They carry this positional information about $\alpha$'s
location by virtue of having the same speed ($\pm 1$).

To see how these elements work together over a space-time region
consider the left side of the left-hand ($\rho_0 < 1/2$) diagram in
\fig{parAB-filtered}(a). Here, $\alpha$ decays into a $\gamma$ and a
$\mu$ particle. The $\mu$ then collides with a $\beta$ before its
companion $\gamma$ (wrapping around the lattice) does. This indicates
that the low-density white region, whose right border is the $\gamma$,
is larger than the black region bordered by the $\mu$.  The $\mu$-$\beta$
collision creates a new particle, a $\delta$, that carries this information
(``low-density domains'') to the left, producing more low-density area.
$\delta$, a fast moving particle, catches up with the $\gamma$ (``low
density'') and annihilates it, producing \domain{0} over the entire lattice.
The result is that the white region takes over the lattice before the
maximum number of iterations has passed. In this way, the classification
of the (low density) IC has been correctly determined by the spatial
algorithm---the steps we have just described. In the case of \caS{par}{a},
this final decision is implemented by $\delta$'s velocity being three
times that of $\gamma$'s.

On the right side of the right-hand ($\rho_0 > 1/2$) diagram in
\fig{parAB-filtered}(a), a converse situation emerges: $\gamma$
collides with $\beta$ before $\mu$ does. The effective decision
indicates that the black region bordered by $\mu$ is larger than the
white region bordered by $\gamma$. In symmetry with the $\mu$-$\beta$
interaction described above, the $\gamma$-$\beta$ interaction creates
the $\eta$ particle that catches up with the $\mu$ and the two
annihilate.  In this way, the larger black region takes over and the
correct density classification is effected.

A third type of particle-based information processing is illustrated
at the top left of the right-hand diagram in
\fig{parAB-filtered}(a). Here, an $\alpha$ decays
into a $\gamma$ and a $\mu$. In this case, the white region bordered
by $\gamma$ is smaller than the black region bordered by $\mu$. As
before, $\gamma$ collides with the $\beta$ on its left, producing
$\eta$. However, there is another $\beta$ particle to the right of
$\mu$. Instead of the $\mu$ proceeding on to eventually collide with
the $\eta$, the $\mu$ first collides with the second $\beta$. Since
the $\mu$ borders the larger of the two competing regions, its
collision is slightly later than the $\gamma$-$\beta$ collision to its
left. The $\mu$-$\beta$ collision produces a $\delta$ particle
propagating to the left. Now the $\eta$ and the $\delta$ approach
each other at equal and opposite speeds and collide. Since $\eta$ is
carrying the information that the white region should win and $\delta$
is carrying the information that the black region should win, their
collision appropriately results in an ``ambiguity'' signal---here, a
$\beta$ that later on interacts with particles from greater distances.
But since $\eta$ traveled farther than $\delta$ before their collision,
a $\beta$ is produced that is is shifted to the right from the
original $\alpha$. The net effect---the net geometric subroutine---is
to shift the location of density ambiguity from that of the original
$\alpha$ particle in the IC to a $\beta$ moved to the right a distance
proportional to the large black region's size relative to the white
region's size.

%%%%%%%%%%%%%%%%%%% PAR_AB-HIGH-FIL %%%%%%%%%%%%%%%%%%%  
\begin{figure}
\centerline{\psfig{figure=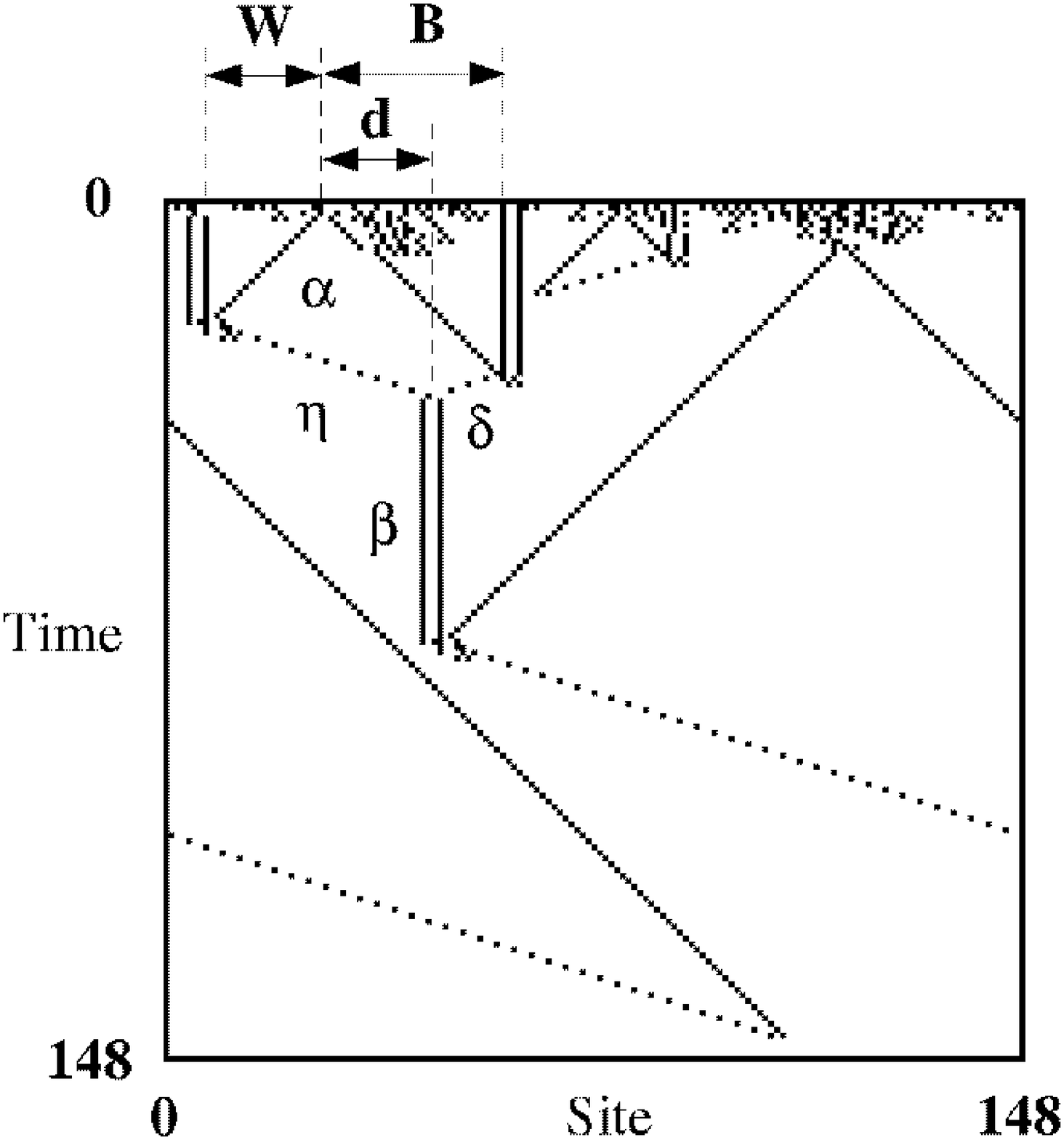,height=4in}}
\caption{An enlargement and relabeling of the right diagram of
  \fig{parAB-filtered}(a) with some particle labels omitted for clarity.
  $\bf W$ is the length of the leftmost white region, $\bf B$ is the
  length of the black region to its right, and $\bf d$
  is the amount by which the $\beta$ produced by the $\eta$--$\delta$
  interaction has been shifted from the leftmost $\alpha$.  Given the
  particle velocities listed in Table~\ref{particle-table_a} and using
  simply geometry, it is easy to calculate that
  ${\bf d} = 2({\bf B}-{\bf W})$.
\label{shifted-beta}}
\end{figure}
%%%%%%%%%%%%%%%%%%%%%%%%%%%%%%%%%%%%%%%%%%%%%%%%%%%%%%%%

Even though this $\beta$ encodes ambiguity that cannot be resolved by the
information currently at hand---that is, the information carried by the
$\eta$ and $\delta$ that produce it---this $\beta$ actually carries
important information in its location, which is shifted to the right
from the original $\alpha$. To see this, refer to \fig{shifted-beta},
an enlargement of the right diagram of \fig{parAB-filtered}(a) with some
particle labels omitted for clarity. $\bf W$ and $\bf B$ denote the
lengths of the indicated white (low density) and black (high density)
regions in the IC. Given the particle velocities listed in
Table~\ref{particle-table_a} and using simple geometry it is easy to
calculate that the $\beta$, produced by the $\eta$-$\delta$ interaction,
is shifted to the right by $2({\bf B}-{\bf W})$ cells from the
$\alpha$'s original position. The shift to the right means that the
high-density region (to the left of the leftmost $\beta$) has gained
${\bf B}-{\bf W}$ sites in size as a result of this series of
interactions. In terms of relative position the local particle
configuration $\beta\alpha\beta$ becomes $\beta\gamma\mu\beta$
and then $\eta\delta$, which annihilate to produce a final $\beta$. This
information is used later on, when the rightmost $\gamma$ collides
with the new $\beta$ before its partner $\mu$ does, eventually leading
to black taking over the lattice, correctly classifying the ($\rho_0 >
1/2$) IC.

It should now be clear in what sense we say that particles store and
transmit information and that particle collisions are the loci of decision
making. We described in detail only two such scenarios. As can be seen from
the figures, this type of particle-based information processing occurs
in a distributed, parallel fashion over a wide range of spatial and temporal 
scales. The functional organization of the information processing can be
usefully analyzed at three levels: (i) the information stored in the
particles and decisions made during their interaction, (ii) geometric
subroutines that are coordinated groupings of particles and interactions
that effect intermediate-scale functions, and (iii) the net spatial
computation over the whole lattice and from $t=0$ to
$t=T_{\textsub{max}}$.

In the next section we will argue that these levels of description of
a CA's computational behavior---in terms of information transmission
and processing by particles and their interactions---is analogous to,
but significantly extends, Marr's ``representation and algorithm''
level of information processing. It turns out to be the most useful
level for understanding and predicting the computational behavior of
CAs, both for an individual CA operating on particular ICs and also
for understanding how the GA evolved the progressive innovations in
computational strategies over succeeding generations. (We will put
these latter claims on a quantitative basis shortly.)

%%%%%%%%%%%%%%%%%%% PAR_A-ERRORS %%%%%%%%%%%%%%%%%%%  

\begin{figure}
\centerline{\psfig{figure=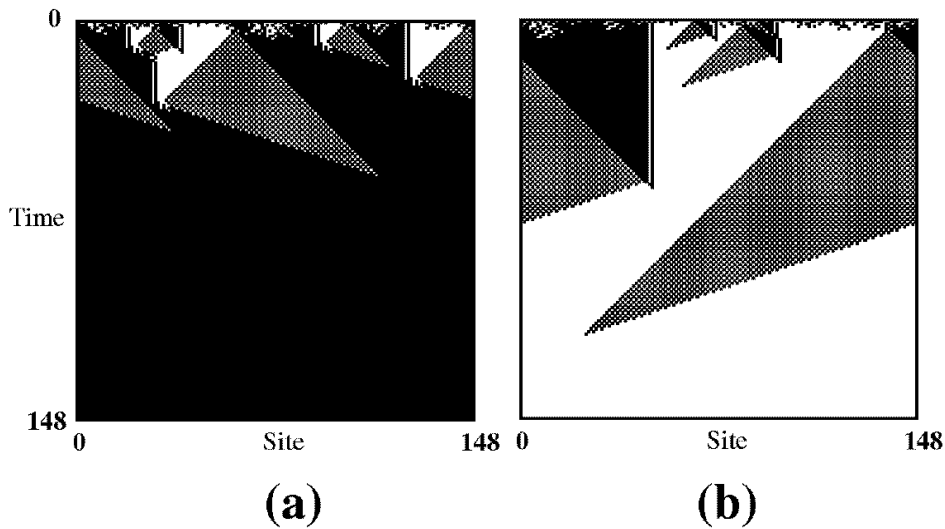,height=3in}}
\caption{(a) Type-1 misclassification by \caS{par}{a}, with $\rho_0 = 0.48$.
  Even though $\rho_0 < \rho_c$, at $t_c$ (here, $t=8$) the lengths of
  the black regions sum to 65 cells and the lengths of the white
  regions sum to 44 cells. This leads to a misclassification of IC
  density. (b) Type-2 misclassification by \caS{par}{a}, starting with
  $\rho_0 = 0.52$. At $t_c$ (here also, $t=8$) the sum of the lengths
  of the black regions is 65 cells and the sum of the lengths of the
  white regions is 61 cells.  However, even though these condensed
  lengths correctly reflect the fact that $\rho_0 > \rho_c$, the black
  regions in the IC's center occur within white regions in such a way
  that they get cut off.  Ultimately this yields a large white region
  that wins over the large black region and the IC is misclassified.
\label{parA-errors}}
\end{figure}

% ICs for this figure:
% (a) 11010100110110111011100111011111000111111101101010101100101000000111101011000011000110000111011000100010110001100100011101000010100101110001111000110
% (b) 10010011101010111110111111000101010000001001010010001100000010001111011111111101011111010100111101100010111011110110111101001001100101110101000000010
%%%%%%%%%%%%%%%%%%%%%%%%%%%%%%%%%%%%%%%%%%%%%%%%%%%%%%%%%%%%%%

\caS{par}{a} almost always iterates to either all $0$s or all $1$s
within $T_{\textsub{max}} = 2N$ time steps.
The errors it makes are almost always due to
the wrong classification being reached rather than no classification
being effected.  \caS{par}{a} makes two types of misclassifications.
In the first type, illustrated in \fig{parA-errors}(a), \caS{par}{a}
reaches the condensation time $t_c$ having produced a configuration
whose density is on the opposite side of $\rho_c$ than was $\rho_0$.
The particles and interactions then lead, via a correct geometric
computation, to an incorrect final configuration. In the second type
of error, illustrated in \fig{parA-errors}(b), the density
$\rho_{t_c}$ is on the same side of the threshold as $\rho_0$, but the
configuration is such that islands of black (or white) cells are
isolated from other black (or white) regions and get cut off. This
error in the geometric computation eventually leads to an incorrect
final configuration. As $N$ increases, this type of error becomes
increasingly frequent and results in the decreasing \perf{10^4}{N}
values at larger $N$; see Table~\ref{ca-table}.

%%%%%%%% COMPUTATIONAL STRATEGY OF PAR_B %%%%%%%%%%%%

\subsection{Computational Strategy of \caS{par}{b}---Failure Analysis}

As noted in Table~\ref{particle-table_b}, the space-time behavior of
\caS{par}{b} exhibits three regular domains: \domain{0} (white),
\domain{1} (black), and \domain{2} (striped). The size-competition
strategy of \caS{par}{b} is similar to that of \caS{par}{a}. In
\caS{par}{b}, the striped region plays the role of \caS{par}{a}'s
checkerboard domain. However, when compared to \caS{par}{a}, the roles
of the two domain boundaries \domain{0}\domain{1} and
\domain{1}\domain{0} are now reversed. In \caS{par}{b},
\domain{0}\domain{1} is stable, while \domain{1}\domain{0} is unstable
and decays into two particles. Thus, the strategy used by \caS{par}{b}
is, roughly speaking, a $0$-$1$ site-value exchange applied to
\caS{par}{a}'s strategy.  Particles $\alpha$, $\beta$, $\gamma$,
$\delta$, $\eta$, and $\mu$ are all analogous in the two CAs, as are
their interactions, if we exclude three-particle interactions; cf.
Tables \ref{particle-table_a} and \ref{particle-table_b}. They
implement competition between adjacent large white and black regions.
In analogy with the preceding analysis for \caS{par}{a}'s strategy,
these local competitions are decided by which particle, a $\gamma$ or
a $\mu$, reaches a $\beta$ first.

%%%%%%%%%%%%%%%%%%% PAR_B-ERRORS %%%%%%%%%%%%%%%%%%%  

\begin{figure}
\centerline{\psfig{figure=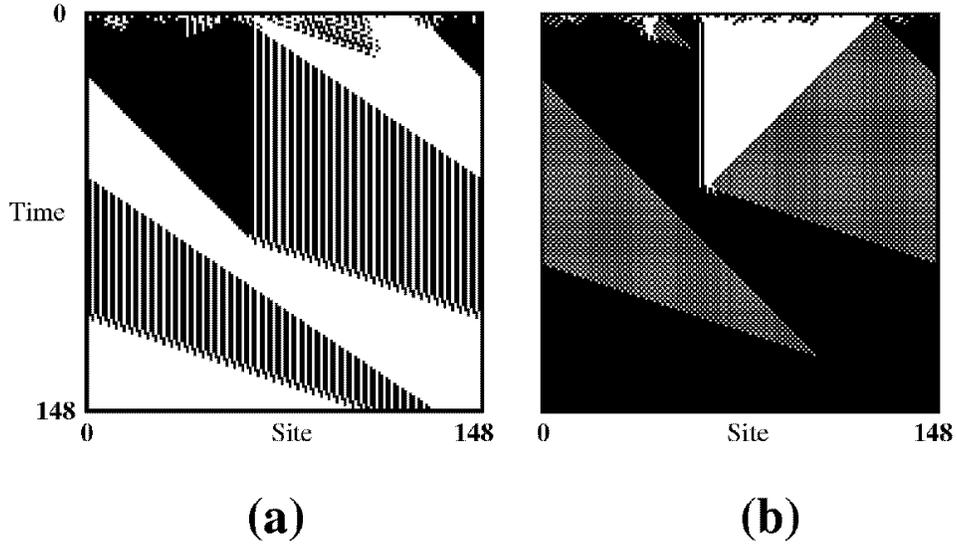,height=3in}}
\caption{(a) Misclassification by \caS{par}{b}, with  
  $\rho_0 = 0.52$.  (By $T_{\textsub{max}} = 2N$, the CA reaches an
  all-$0$s fixed point.) (b) Correct classification by \caS{par}{a} on the
  same IC.
\label{parB_errors}}
\end{figure}

% IC for this figure:
% 01101110010110110111110111101101110000101100101110110111101000100110010110011000101001010100001101001011000001001000000000100011011000111110111111011  

%%%%%%%%%%%%%%%%%%%%%%%%%%%%%%%%%%%%%%%%%%%%%%%%%%%%%%%%%%%%%%

In \caS{par}{a}, $\gamma$ and $\mu$ each approach $\beta$ at the rate of one
cell per time step. In \caS{par}{b}, although $\gamma$ is now a stationary
particle, it also effectively approaches $\beta$ at the rate of $1$
cell per time step, since $\beta$ moves with velocity $1$. $\mu$
approaches $\beta$ at the rate of $1/2$ cell per time step, the
velocity of $\mu$ minus the velocity of $\beta$. Thus, there is an
asymmetry in \caS{par}{b}'s geometric computation that can result in
errors of the type illustrated in \fig{parB_errors}(a).  There the IC, with
$\rho = 0.52$, condenses around iteration $t_c \approx 20$ into a block
of 85 black cells adjacent to a block of 64 white cells. The $\gamma$
particle, traveling at velocity 1 relative to $\beta$, reaches $\beta$
in approximately 85 time steps.  The $\mu$ particle, traveling at
velocity $1/2$ relative to $\beta$, reaches $\beta$ in approximately
103 time steps.  Thus, even though the black cells initially outnumber
the white cells, the black region is cut off first and white
eventually wins out, yielding an incorrect classification at time step
165. In contrast, \caS{par}{a}, with its symmetric particle velocities,
reaches a correct classification on this same IC (\fig{parB_errors}(b)).

%%%%%%%%%%%%%%%%%%% PAR_B-ASYMMETRIES %%%%%%%%%%%%%%%%%%%  

\begin{figure}
\centerline{\psfig{figure=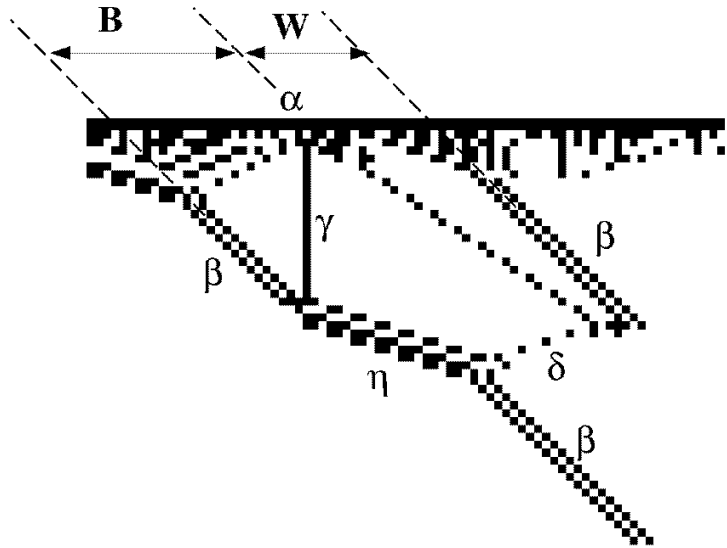,height=4in}}
\caption{Blow-up of part of right-hand diagram of \fig{parAB-filtered}b, 
illustrating asymmetries in \caS{par}{b}'s particle velocities that can result
in misclassifications.
\label{parB-asymmetries}}
\end{figure}

%%%%%%%%%%%%%%%%%%%%%%%%%%%%%%%%%%%%%%%%%%%%%%%%%%%%%%%%%%%%%%

Like \caS{par}{a}, \caS{par}{b} makes two types of classification
errors---the type in which \caS{par}{b} reaches $t_c$ with a configuration
whose density $\rho_{t_c}$ is on the opposite side of $\rho_c$ than
$\rho_0$ and the type illustrated in \fig{parB_errors}(a). The
first type (``type 1'') is an error in how the CA condenses into
domains and particles. The second type (``type 2'') is due to
asymmetries in particle velocities. Consider \fig{parB-asymmetries}, a
blow-up of part of the right-hand diagram of \fig{parAB-filtered}(b).
To the left of the $\alpha$ (labeled) is an isolated black island and to the
right is a white island. Together these two contiguous islands are bounded by
two $\beta$ particles on either side. Inside, as in \caS{par}{a}, an $\alpha$
decays into a $\gamma$ and a $\mu$.  The resulting set of local
particle interactions is such that the two islands compete for space
within the two bounding $\beta$'s, ending with the creation of a new $\beta$.
If $\bf W$ and $\bf B$ respectively denote the lengths of the white and black
islands, then after a series of interactions---$\beta\gamma\mu\beta
\rightarrow \eta\delta \rightarrow \beta$---the white region (to the left
of the original leftmost $\beta$) gains $2{\bf W} - {\bf B}$ sites in size. Thus,
to increase this region's size the internal white island must be at least
half the size of its adjacent black island.

It is evident, therefore, that unlike
\caS{par}{a}, there are asymmetries in \caS{par}{b}'s particle
``logic'', and these are biased in favor of classifying high
densities.  These asymmetries are what make
\performance{10^4}{N}{\caS{par}{b}} lower than
\performance{10^4}{N}{\caS{par}{a}}. (See Table \ref{ca-table}.)

%%%%%%%%%%% Significance of the Particle-Level Description %%%%%%%%%%%

\section{Significance of the Particle-Level Description}

There are several alternative ways in which cellular automata such as
\caS{par}{a} and \caS{par}{b} can be described as performing a
computation. Marr anticipated some of these in delineating the various
levels of information processing in vision \cite{Marr82}. In
principle, our CAs are completely described by the 128 bits in their
look-up tables.  This is too low-level a description, however, to be
useful for understanding how a given CA performs the $\rho_c=1/2$
task.  Using this level is like trying to understand how a pocket calculator
computes the square root function by examining the physical equations
of motion for the electrons and holes in the calculator's silicon circuitry.

Moreover, attempting interpretation at this level also violates, in a
sense, one of the central tenets of the century-long study of dynamical
systems, namely, that for nonlinear systems (e.g., most CAs), the local
space-time equations of motion do not {\em directly}
determine the system's long-term behavior. In the case of CAs it is not
the individual look-up table neighborhood-output-bit entries acting over
a single time step that directly give rise to the observed computational
strategy. Instead, it is the interaction of {\em subsets} of CA look-up
table entries that over a {\em number of iterations} leads to the
emergence of domains, particles, and interactions.

A second possibility for describing computational behavior in CAs is in
terms of its detailed space-time behavior---i.e., the series of raw
configurations of $0$s and $1$s. Again, this description is too low level for
understanding how the solutions to the task are implemented. This
approach is like trying to understand how a calculator's square root
function is performed by taking a long series snapshots of the positions
and velocities of the electrons and holes traveling through the
integrated circuits. This prosaic view is analogous to Marr's
``hardware implementation'' level of description \cite{Marr82}.  

A third possibility is to describe the CA in terms related to the
task's required input/output mapping and the task's computational
complexity.  For example, on a particular set of $10^4$ random ICs,
half with $\rho < 1/2$ and half with $\rho > 1/2$, \caS{par}{a}
correctly classified 81\% of the $\rho < 1/2$ ICs and 74\% of the
$\rho > 1/2$ ICs.  On average \caS{par}{a} took 81 time steps to reach
a fixed point; the maximum time was 227.  The computational complexity
of the $\rho_c = 1/2$ task on a serial architecture is ${\cal O}(N)$.
This kind of operational analysis is roughly at Marr's ``computational
theory'' level.

None of these levels of description gives much insight into {\it how}
the task is being performed by a particular CA in terms of what
information processing is being done and how it leads to a particular
measured performance. What is needed is an intermediate-level
description whose primitives are informationally related to the task
at hand. This is what the computational mechanics level of particles
and particle interactions gives us. How ever one might detect the
primitives at this level, it is analogous to Marr's ``representation
and algorithm'' level, in which particles can be seen as representing
aspects of the IC and their actions and interactions can be seen as
the CA's emergent algorithm.

Representations, in the form of data structures, and
algorithms have been studied extensively for von Neumann-style
computers, but there have been few attempts to define such notions for
decentralized spatially extended systems such as CAs.  One can, of
course, in principle implement any standard data structure and
algorithm in a computation-universal CA, such as the game of Life
CA \cite{BerlekampEtAl82}, by simulating a von Neumann-style computer.
However, this is not a particularly useful notion of information
processing if one's goals are to understand how nonlinear systems in
nature compute. It is even more problematic if one wishes to design
computation in complex decentralized spatially extended architectures.
We believe that it will be essential to develop new
``macroscopic-level'' vocabularies in order to explain how collective
information processing takes place in such architectures. (One benefit
of this development would be an understanding of how to program these
architectures in genuinely parallel ways.)

A close reading shows that Marr's analysis of the descriptional levels
required for visual processing misses several key issues. These are
(i) the fact that representations emerge from the dynamics (i.e., are
{\it intrinsic} to the dynamics), (ii) a clear formal definition is
required to remove the subjectivity of detecting these intrinsic
representations, and (iii) their functionality is entailed by a new
level of dynamics, also intrinsic, that describes their interactions.
As illustrated above in several cases, the computational mechanics
framework that we are employing here makes these distinctions and
provides the necessary concepts and methods to address these issues
\cite{Crut92c,Crut93g}. The result is that we can analyze in detail
the emergent computational strategies in the evolved CAs.

Our particle-level description forms an explanatory vocabulary for
emergent computation in the context of one-dimensional, binary-state
CAs. As was described above, particles represent various kinds of
information about the IC and particle interactions are the loci of
decision making that use this information. The resulting particle
``logic'' gives a functional description of how the computation takes
place that is neither directly available from the CA look-up table nor
from the raw space-time configurations produced by iterating the CA.
It gives us a formal notion of ``strategy'', allowing us to see, for
example, how the strategies of \caS{par}{a} and \caS{par}{b} are
similar and how they differ.  One immediate consequence of this level
of analysis is that we can say why \caS{par}{b}'s strategy is weaker.

The level of particles and interactions is not only a qualitative
description of spatial information processing, it also enables us
to make quantitative predictions about computational performance.
In Refs. \cite{CrutchfieldEtAl97a} and \cite{HordijkEtAl98}, 
we describe how to model a CA using its
particle catalog and statistical properties at the condensation time.
For each of several different CAs $\phi$, we compare the model's
prediction for \performance{I}{N}{$\phi$} as well as for the average
time taken to reach a fixed point with the values measured for the
actual $\phi$. Some of these comparisons will be summarized in
Sec.~\ref{particle-models}. The degree to which a model's predictions
agree with the corresponding CA's behavior indicates the degree to
which the particle-level description captures how the CA is actually
performing the computation. Since, as we will show, the model's
predicted performance and the observed performance are very
close, we conclude that the particle-level description accurately
captures the intrinsic computational capability of the evolved CAs.

%%%%%%%%%%%%%%%%%% EVOLUTIONARY HISTORY %%%%%%%%%%%%%%%%%%%%%%%%%

\section{Evolutionary History of \caS{par}{a}: Innovation, Contingency,
and Exaptation \label{evolutionary-history}}

%%%%%%%%%%%%%%%%%%% PAR_A-FAMILY-TREE %%%%%%%%%%%%%%%%%  

\begin{figure}
\centerline{\psfig{figure=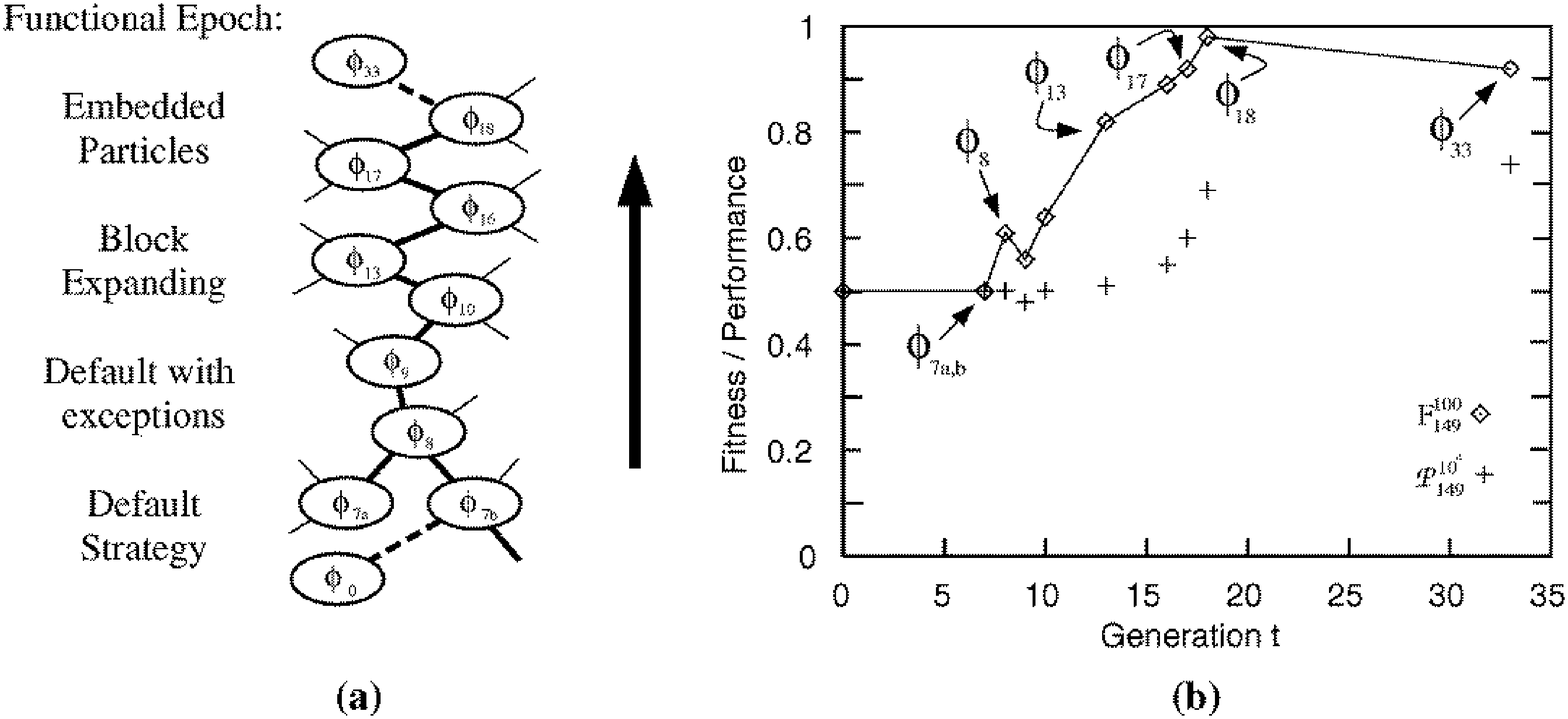,height=4in}}
\caption{(a) Partial ancestral tree of \caS{par}{a}.
  (b) \fitness{100}{149}{$\phi$} (diamonds) and
  \performance{10^4}{149}{$\phi$} (crosses) of the CAs $\phi_i$
  in (a). Six of the data points are marked with the name of the
  corresponding CA.
\label{parA-family-tree}}
\end{figure}

%%%%%%%%%%%%%%%%%%%%%%%%%%%%%%%%%%%%%%%%%%%%%%%%%%%%%%%%%%%%%%

The structural analysis of CA space-time information processing that
we have just outlined also allows us to understand the evolutionary stages
during which the GA produces CAs. Here we will show how the functional
components---domains, particles, and interactions---arise and are
inherited across the evolutionary history of a GA run. We will also
demonstrate a number of evolutionary dynamical phenomena, such as the
historical contingency of functional emergence and the appearance of
initially nonfunctional behaviors that later are key to the final appearance
of high performance CAs.

%%%%%%%%%%%%%%%%%%% EVHISTORY-TABLE %%%%%%%%%%%%%%%%%%%  

%%%% I fixed these to be the values given in Figure 13(b). -- Melanie 

\begin{table}
\begin{center}
\begin{tabular}{|c|c|c|c|}
\hline
CA Name  &  Rule Table (Hexadecimal) &  \fit{100}{149} & \perf{10^4}{149} \\ \cline{1-4}
\ca{7b}        & {\tt F6EFFFFFFFFFFFFF} & 0.50  & 0.500 \\
               & {\tt 6B9F7F93FFFFBFFF} &       &       \\ \cline{1-4}
\ca{8}         & {\tt 0400448102000FFF} & 0.61  & 0.500 \\
               & {\tt 6B9F7F93FFFFBFFF} &       &       \\ \cline{1-4}
\ca{13}        & {\tt 0400458100000FFF} & 0.82  & 0.513 \\ 
               & {\tt 6B9F77937DFFFF7F} &       &       \\ \cline{1-4}
\ca{17}        & {\tt 0500458100000FBF} & 0.92  & 0.595 \\ 
               & {\tt 6B9F75937FBFFF5F} &       &       \\ \cline{1-4}
\ca{18}        & {\tt 0500458100000FBF} & 0.98  & 0.691 \\ 
               & {\tt 6B9F75937FBDF77F} &       &       \\ \cline{1-4}
\ca{33}        & {\tt 0504058100840FB7} & 0.97  & 0.735 \\ 
               & {\tt 4BBF55837FBDF77F} &       &       \\ \cline{1-4}
\end{tabular} 
\end{center}
\caption{CA chromosomes (look-up table output bits) given in
hexadecimal, \fit{100}{149}, and \perf{10^4}{149} for the six
ancestors of \caS{par}{a} described in this section. (See
\fig{ca-table} for directions on how to recover the CA rule
table outputs from the hexadecimal code.) The \fit{100}{149}
values in this table are those calculated in the CA's generation of
birth by the GA; the \perf{10^4}{149} values given are
the means over 100 trials of the performance function, 
calculated after the run was complete. When tested over 100 trials, 
the standard deviation of \fit{100}{149} is
approximately $0.02$ 
and the standard deviation of \perf{10^4}{149} 
is approximately $0.005$ for each CA. 
\label{evhistory-table}}
\end{table}

%%%%%%%%%%%%%%%%%%%%%%%%%%%%%%%%%%%%%%%%%%%%%%%%%%%% 

To begin, \figs{parA-family-tree}(a) and \ref{parA-family-tree}(b) illustrate \caS{par}{a}'s
evolutionary history. \fig{parA-family-tree}(a) gives a partial tree
of the parent-child relationships between some of \caS{par}{a}'s
ancestors, each numbered by its generation of birth. Note that, since
elite CAs can survive for more than one generation, parents and offspring,
e.g. \ca{10} and \ca{13}, can have nonconsecutive generation labels.
The CAs listed are those with the best fitness in the generation in
which they arose.  Table~\ref{evhistory-table} lists the look-up
tables, \fit{100}{149}, and \perf{10^4}{149} for the six ancestors
of \caS{par}{a} described below.

\fig{parA-family-tree}(b) plots \fit{100}{149} (diamonds) and
\perf{10^4}{149} (crosses) versus generation of birth for each of
these ancestors.  In generations 0--7 the best CA in the population
has \fit{100}{149} $=$ \perf{10^4}{149} $= 0.5$, achieved by a
``default'' strategy like those of \fig{defAB}.  Starting at
generation 8, evolution proceeds in a series of abrupt increases in
\fit{100}{149}. More gradual increases are seen in
\perf{10^4}{149}; of course, this statistic is not available
to and thus is not used by the GA. The occasional small decreases result
from the stochastic nature of the fitness and performance evaluations. 

\begin{figure}
\centerline{\psfig{figure=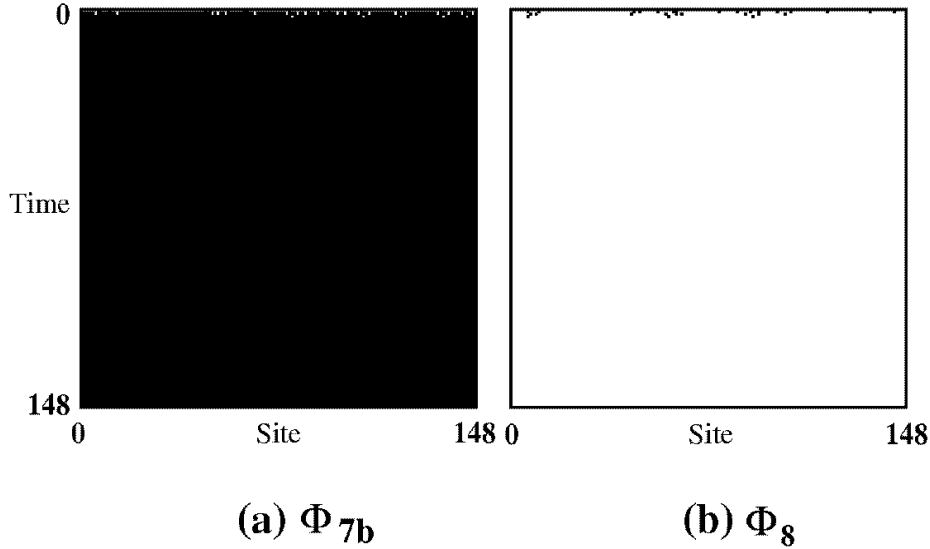,height=3in}}
\caption{Space-time behavior of generation 7 and 8 ancestors,
$\phi_{7b}$ and $\phi_8$, of \caS{par}{a}. Both start from the same IC
with $\rho_0 = 0.11$. 
\label{7-8}}
\end{figure}

\begin{figure}
\centerline{\psfig{figure=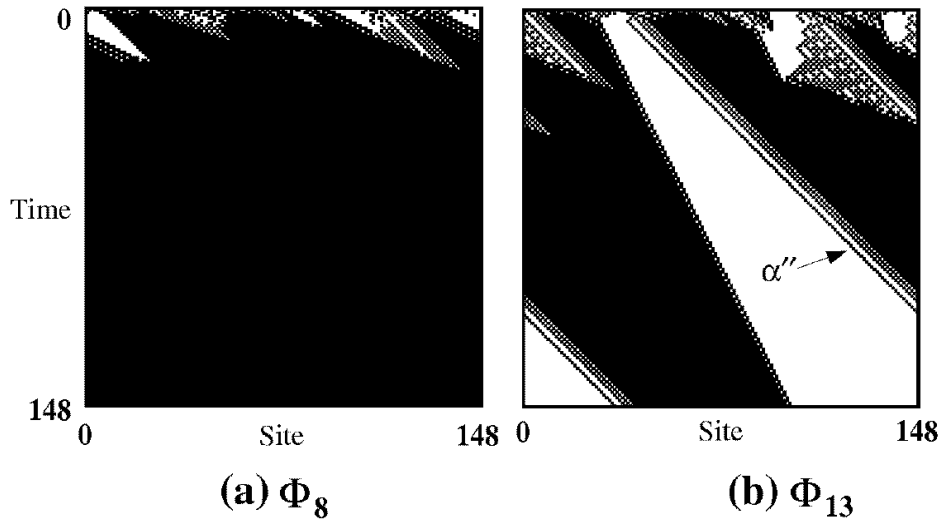,height=3in}}
\caption{
  Space-time behavior of generation 8 and 13 ancestors, $\phi_8$ and
  $\phi_{13}$, of \caS{par}{a}. Both start from the same IC with
  $\rho_0 = 0.47$. (The significance of $\alpha^{\prime\prime}$ is
  explained below, when \ca{13} and \ca{17} are compared.)
  \label{8-13}
  }
\end{figure}

The goal now is to use the functional analysis to understand why these
increases come about.  To do so, we present a series of space-time
diagrams, in Figs. \ref{7-8}-\ref{33-64}, that compare space-time
behaviors of CAs along the ancestral tree of \fig{parA-family-tree}(a).
In each figure, space-time behavior with the same IC is given for two
ancestrally related CAs to highlight the similarities and
evolutionary innovations.

%%%%%%%%%%%%%%% COMPARING PAIRS OF ANCESTORS %%%%%%%%%%%%%%%%%%%

%%%%%%%%%%%%%%%%%%% 7-8 %%%%%%%%%%%%%%%%%  

%%%%%%%%%%%%%%%%%%%%%%%%%%%%%%%%%%%%%%%%%%

\subsection{\ca{7b} and \ca{8} (\fig{7-8})}

Here the IC has very low density: $\rho_0 = 0.11$.  \ca{7b}, a
``default'' CA, always iterates to all $1$s, and in \fig{7-8}(a)
misclassifies the IC. \ca{7a} (not shown) is a default CA that always
iterates to all 0s.  \ca{7b}'s look-up table contains mostly 1s (see
Table~\ref{evhistory-table}) and \ca{7a}'s look-up table contains
mostly 0s.  They crossed over at locus 52 to produce \ca{8}: therefore
the first part of \ca{8}'s look-up table contains mostly $0$s, and the
rest is mostly $1$s.  In \fig{7-8}(b) \ca{8} iterates to all $0$s and
correctly classifies the IC.

%%%%%%%%%%%%%%%%%%% 8-13 %%%%%%%%%%%%%%%%%  

%%%%%%%%%%%%%%%%%%%%%%%%%%%%%%%%%%%%%%%%%%%%%%%%%%%%%%%%%%%%%%

\subsection{\ca{8} and \ca{13} (\fig{8-13})}

Here $\rho_0 = 0.47$.  In \fig{8-13}(a) \ca{8} quickly iterates to all
$1$s.  This is its more typical behavior than that shown in
\fig{7-8}(b); very small regions of black quickly grow to take over
the entire lattice. In this way, \ca{8} is only slightly better than
a default CA like \ca{7b}; it correctly classifies all
high-density ICs and only a small number of very low density ICs. Note
that while \fitness{100}{149}{\ca{8}} $> 0.5$,
\performance{10^4}{149}{\ca{8}} remains at $0.5$.  \ca{8} can be said
to be carrying out a ``default-with-exceptions'' strategy. All runs that
produced such strategies went on to converge on either block-expanding
strategies or embedded-particle strategies.

Interestingly, the checkerboard domain $\Lambda^2 = \{(01)^+\}$ is
produced by \ca{8} on some ICs (\fig{8-13}). However, $\Lambda^2$ does
not contribute to \ca{8}'s fitness or performance.
It is a functionally neutral
feature.  To determine this, we modified \ca{8}'s rule table to
prevent the checkerboard domain from propagating.  The two relevant
entries are $0101010 \rightarrow 0$ and $1010101 \rightarrow 1$.
Flipping the output bit on either or both of these entries produces
CAs with \fit{100}{149} $=0.61$ and \perf{10^4}{149} $= 0.5$; that is,
fitness and performance identical to those of \ca{8}. (The standard
deviations of \fit{100}{149} for this and the other variant CAs
discussed in this section were approximately $0.02$.  The standard deviations of
\perf{10^4}{149} were approximately $0.005$.) Appropriating biological
terminology, we can consider the checkerboard domain, at this generation,
to be an adaptively neutral trait of \ca{8}.

\ca{13} represents a steep jump in fitness over \ca{8}, as seen in
\fig{parA-family-tree}(b). \ca{13} is a block-expanding CA. It maps ICs
to all $1$s unless there is a sufficiently large block of adjacent
$0$s in the IC, in which case that block expands to eventually fill up
the entire lattice, as in \fig{8-13}(b), which is a correct
classification by $t = T_{\textsub{max}}$. On some ICs, \ca{13} also
produces a checkerboard domain and a similar but less ordered region;
the latter can be seen in \fig{8-13}(b). We determined, in a fashion
similar to that just explained above, that these traits also were
adaptively neutral.

%%%%%%%%%%%%%%%%%%% 13-17 %%%%%%%%%%%%%%%%%  

\begin{figure}
\centerline{\psfig{figure=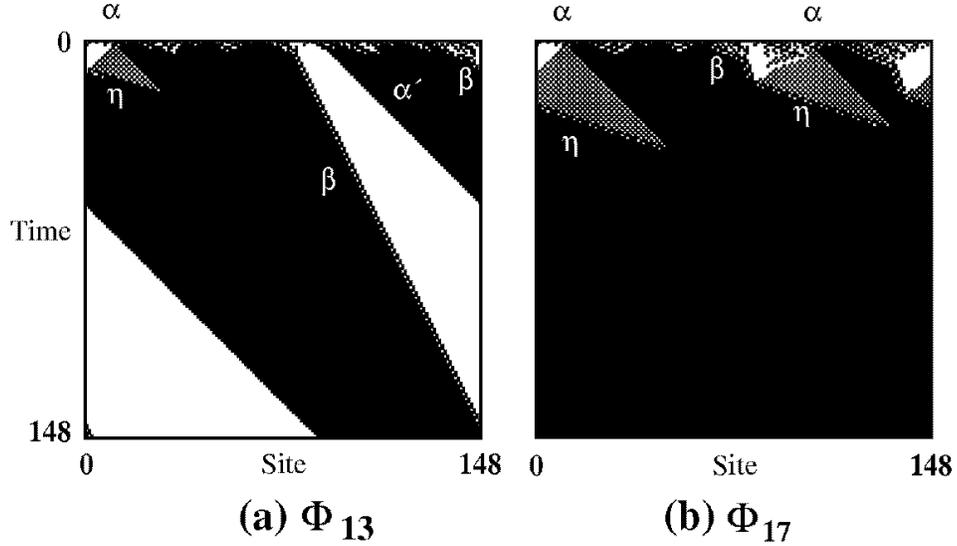,height=3in}}
\caption{Space-time behavior of \caS{par}{a} ancestors, $\phi_{13}$ and
$\phi_{17}$ arising in generation $13$ and $17$.
Both start from the same IC with $\rho_0 = 0.58$.  
\label{13-17}}
\end{figure}

%%%%%%%%%%%%%%%%%%%%%%%%%%%%%%%%%%%%%%%%%%%%%%%%%%%%%%%%%%%%%%

\subsection{\ca{13} and \ca{17} (\fig{13-17})}

Here $\rho_0 = 0.58$.  \ca{13} expands blocks of $0$s on many ICs with
$\rho > 1/2$, including the one in this figure, resulting in
misclassifications.  In fact, many high-density ICs with 
$\rho \approx 0.5$ are misclassified and, while \ca{13} has markedly
higher \fit{100}{149} than its ancestors, its performance \perf{10^4}{149} is
only marginally improved (see Table~\ref{evhistory-table}). 

\ca{13} creates three types of boundaries between white and black
domains. Two of them are shown in \fig{13-17}(a), labeled $\alpha$ and
$\alpha^\prime$. The $\alpha$, like \caS{par}{a}'s $\alpha$, exists
for only a single time step and then decays into $\eta$ and $\mu$,
whereas $\alpha^\prime$ remains stable. A third type, $\alpha''$, does
not appear for this IC but can be seen in \fig{8-13}(b). $\alpha'$ and
$\alpha^{\prime\prime}$ support the block-expanding strategy, whereas
$\alpha$ leads to a competition between white and black regions
similar to that seen in \caS{par}{a}.

In contrast, consider \fig{13-17}(b), where the same $\rho=0.58$ IC is
correctly classified by \ca{17}. Recalling Table~\ref{evhistory-table}
we see that \ca{17}'s \fit{100}{149} and \perf{10^4}{149} are both
substantially higher than those of \ca{13}. \ca{17}'s higher
\fit{100}{149} and \perf{10^4}{149} can be explained at the particle
level. The particles are labeled in \fig{13-17}(a) and \fig{13-17}(b).

\ca{17} creates the same set of particles as \ca{13} (on some ICs it
expands $0$-blocks, not shown in \fig{13-17}(b)) but with different
frequencies of occurrence: $\alpha'$ and $\alpha''$ appear less often
than in \ca{13} and $\alpha$ appears more often. Thus, \ca{13} is more
likely to expand $0$-blocks, and thus make more errors, than \ca{17}
on ICs for which $\rho_0 > 0.5$.  Given $100$ randomly generated $\rho
> 0.5$ ICs, $\alpha'$ and $\alpha''$ were created by \ca{13} in $86$\%
of the ICs and by \ca{17} in $12$\% of the ICs. Whenever $\alpha'$ or
$\alpha''$ are created, the final configuration will be all $0$s
regardless of whether $\alpha$ is created. That is, block expanding
dominates other behaviors. This explains why flipping output bits
to suppress the checkerboard domain does not significantly affect
\ca{13}'s \fit{100}{149} and \perf{10^4}{149}, but does significantly
affect these values for \ca{17}.  When the checkerboard domain was
suppressed in \ca{17}, \fit{100}{149} decreased only to 0.86 but
\perf{10^4}{149} decreased to 0.54.

Following Gould and Vrba \cite{Gould&Vrba82}, we consider the
checkerboard domain $\Lambda^2$ to be an example of an
``exaptation''---a trait that has no adaptive significance when it
first appears, but is later co-opted by evolution to have adaptive
value.  According to Gould and Vrba, such traits are common in
biological evolution. In the evolutionary innovation that goes from
\ca{13} to \ca{17} the exaptation of $\Lambda^2$ in \ca{13} makes just
this transition to functionality associated with a marked increase in
fitness and performance. This, in turn, leads to the change in dominant
computational strategy away from block expanding.

%%%%%%%%%%%%%%%%%%% 17-18 %%%%%%%%%%%%%%%%%  

\begin{figure}
\centerline{\psfig{figure=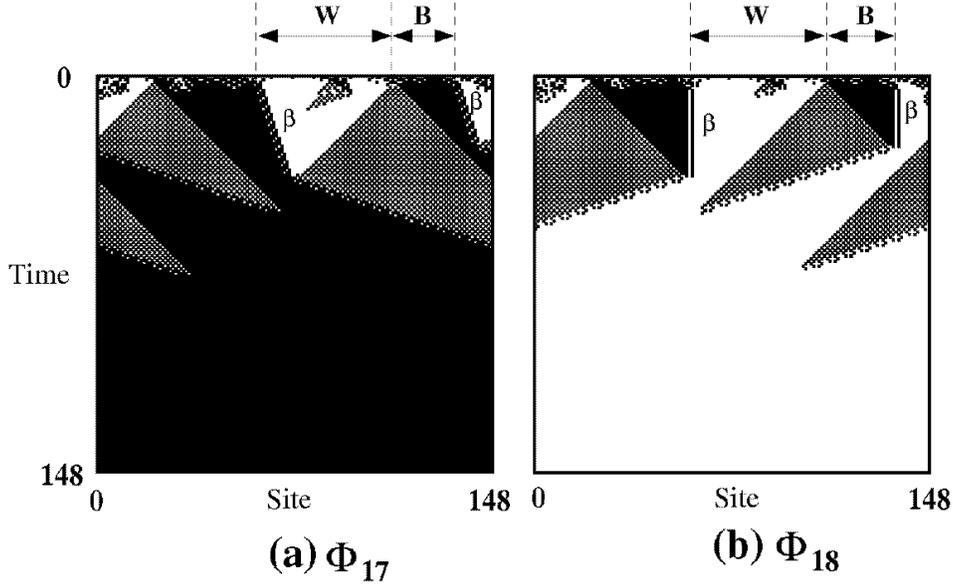,height=3.5in}}
\caption{Space-time behavior of generation 17 and 18 ancestors,
\ca{17} and \ca{18}, of \caS{par}{a}. Both are shown starting from the
same IC that has $\rho_0 = 0.45$. \label{17-18}}
\end{figure}

%%%%%%%%%%%%%%%%%%%%%%%%%%%%%%%%%%%%%%%%%%%%%%%%%%%%%%%%%%%%%%

\subsection{\ca{17} and \ca{18} (\fig{17-18})} 

Here $\rho_0 = 0.45$. The misclassification by \ca{17} (illustrated in
\fig{17-18}(a)) is compared with the correct classification by \ca{17}'s
higher fitness and performance child \ca{18} (\fig{17-18}(b)). Both
CAs create similar particles, but in \ca{17} the velocity of the $\beta$
particle is $1/3$, whereas in \ca{18} its velocity is zero.

In \fig{17-18}(a), the white region (marked $\bf W$) is larger than the
black region to its right (marked $\bf B$). Since the $\beta$ particles
have positive velocity, the black regions to $\bf W$'s left and right both
expand to the right. Coming in from the left, this decreases the size of
$\bf W$. On the other side, the (rightmost) $\beta$ particle moves away
from the $\bf W$ region. This asymmetry allows the $\bf B$ region to
win the size competition, when the $\bf B$ region should not.

The asymmetry between black and white regions is corrected in \ca{18}
by the change in $\beta$'s velocity to zero. This makes the size
competition between black and white regions symmetric. The result,
seen in \fig{17-18}(b), is that the smaller $\bf B$ region is now cut
off by the $\mu$ and $\beta$, the $\bf W$ region is allowed to grow,
and the correct classification is made.

%%%%%%%%%%%%%%%%%%% 18-33 %%%%%%%%%%%%%%%%%  

\begin{figure}
\centerline{\psfig{figure=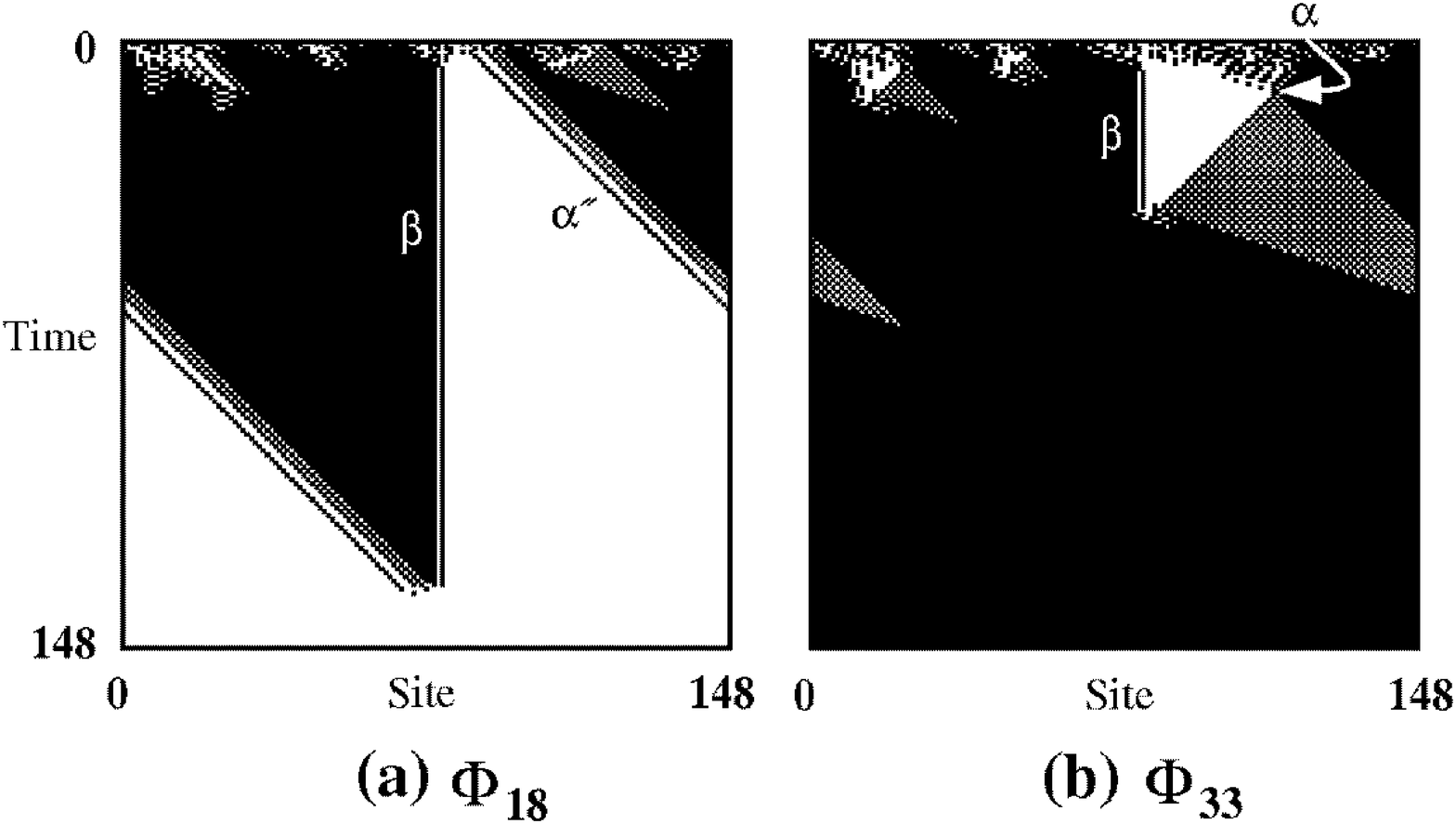,height=3in}}
\caption{Space-time behavior of generation 18 and 33 ancestors,
\ca{18} and \ca{33}, of \caS{par}{a}. Both start from the same IC
with $\rho_0 = 0.61$. 
\label{18-33}}
\end{figure}

%%%%%%%%%%%%%%%%%%%%%%%%%%%%%%%%%%%%%%%%%%%%%%%%%%%%%%%%%%%%%%

\subsection{\ca{18} and \ca{33} (\fig{18-33})}

Here $\rho_0 = 0.61$. \ca{18}, though an improvement over \ca{17}, still 
carries with it remnants of its ancestors' block-expanding past. In
\fig{18-33}(a), \ca{18} misclassifies the IC by creating an $\alpha''$
particle instead of an $\alpha$ particle at a white-black (ambiguous
density) boundary in the IC. Recall that in \ca{17}, $\alpha'$ or
$\alpha''$ particles were created in $12$\% of the random ICs with
$\rho > 1/2$. In \ca{18} this frequency is about the same: $13$\%.
Thus, \ca{18}'s main innovation over \ca{17} is the zero velocity of
the $\beta$ particle and the resulting symmetric size-competition
strategy.

In \ca{33}, a descendant of \ca{18}, neither $\alpha'$ or $\alpha''$
particles are created. This explains the higher \fit{100}{149} and
\perf{10^4}{149} of \ca{33} over those of \ca{18}.

%%%%%%%%%%%%%%%%%%% 33-64 %%%%%%%%%%%%%%%%%  

\begin{figure}
\centerline{\psfig{figure=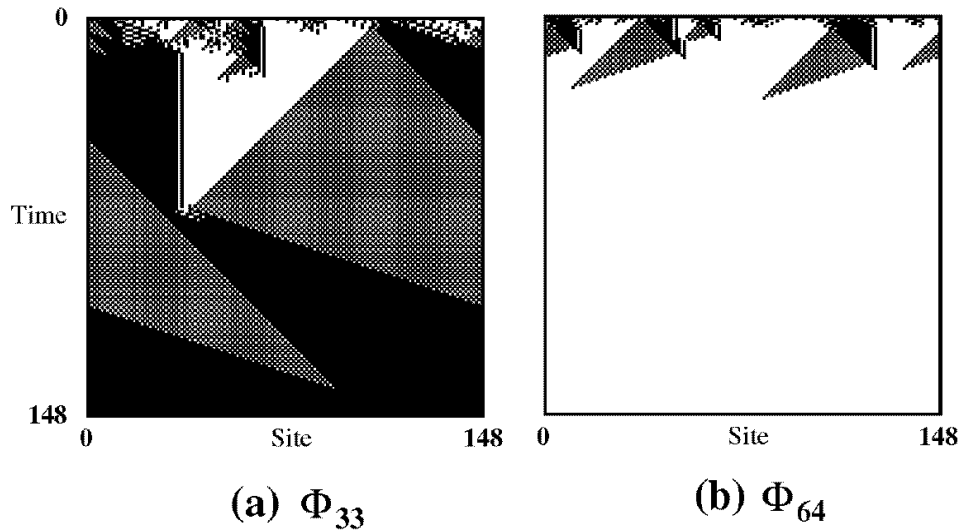,height=3in}}
\caption{Space-time behavior of generation 33 ancestor \ca{33}
  of \caS{par}{a} and \ca{64} (itself \caS{par}{a}). Both start from
the same IC, which has density $\rho_0 = 0.45$. \label{33-64}}
\end{figure}

%%%%%%%%%%%%%%%%%%%%%%%%%%%%%%%%%%%%%%%%%%%%%%%%%%%%%%%%%%%%%%

\subsection{\ca{33} and \ca{64} (\fig{33-64})} 

Here, $\rho_0 = 0.45$.  \caS{par}{a}, here named \ca{64} to denote its
generation of birth, outperforms \ca{33} since it classifies more
low-density ICs correctly. On some low-density ICs like the one used
in \fig{33-64}(a), \ca{33} condenses too much of the IC into black
regions, a type 1 error. These then win the size competition,
resulting in a misclassification. \ca{64} makes type 1 errors on
low-density ICs less often (e.g., as seen in \fig{33-64}(b)), it
correctly classifies the same IC as in \fig{33-64}(a). On the same set
of $10^4$ ICs, 62\% of \ca{33}'s errors were on low-density ICs,
whereas only 43\% of \ca{64}'s errors were on low-density ICs.

%%%%%%%%%% PARTICLE MODELS OF EVOLVED CELLULAR AUTOMATA %%%%%%%%%%

\subsection{Particle Models of Evolved Cellular Automata \label{particle-models}}

The ``natural history'' of \caS{par}{a}'s evolution given above
demonstrates how we can understand the jumps in \fit{100}{149} and
\perf{10^4}{149} in terms of regular domains and
particles---functional components in the CA's dynamical behavior. The
GA's actions can be described at a low level as manipulating bits in
CA rule tables via crossover and mutation, but a better understanding
of the evolutionary process emerges when we describe its actions at
the higher level of manipulating particle types, velocities, and
interactions.  An important component of this viewpoint is that the
particles and their interactions lead to higher fitness.
To test the hypothesis more quantitatively, we ask to
what extent the CAs' observed fitnesses and performances can be
predicted from the particle and interaction properties alone. To
this end, in collaboration with Wim Hordijk,
we have constructed ``ballistic'' particle models of the CAs
\ca{8} through \ca{64}. These models are intended to isolate the
particle-level mechanisms and in so doing allow us to determine how
much of the CA behavior this level captures.

A ballistic particle model ${\cal M}_{\phi}$ of a CA $\phi$ consists
of the catalog of particle types, velocities, interactions, and their
frequencies of occurrence at $t_c$. ${\cal M}_{\phi}$ is ``run'' by
first using the particle frequencies to generate an initial
configuration ${\bf s}_{t_c}$ of particles at the condensation time
and then using the catalog of velocities and interactions to calculate
the initial particles' ballistic trajectories and the products of
subsequent particle interactions. The final configuration is reached
when either all particles have annihilated or when
$T_{\textsub{max}}-t_c$ steps have occurred. This configuration and
the actual time at which it was reached gives us ${\cal M}_{\phi}$'s
prediction of what $\phi$'s classification would be for an IC
corresponding to ${\bf s}_{t_c}$ and the time it would take $\phi$ to
reach it. Particle models and their analysis are described in detail
in Refs. \cite{CrutchfieldEtAl97a} and \cite{HordijkEtAl98}.

%%%%%%%%%%%%%%%%%%%% SPE RESULTS %%%%%%%%%%%%%%%%%%%%%%%%%%%%%% 

\begin{table}[htb]
  \begin{center}
  \begin{tabular}{|c|c|c|}
  \hline
  CA Name &  \perf{10^4}{149}($\phi$) & \perf{10^4}{149}(${\cal M}_{\phi}$) \\ \cline{1-3}
  $\phi_{8}$  &  0.500  & 0.500 \\ \cline{1-3} 
  \hline
  $\phi_{13}$ &  0.513  & 0.524 \\ \cline{1-3} 
  \hline
  $\phi_{17}$ &  0.595  & 0.601 \\ \cline{1-3} 
  \hline
  $\phi_{18}$ &  0.691  & 0.747 \\ \cline{1-3} 
  \hline
  $\phi_{33}$ &  0.735  & 0.765 \\ \cline{1-3} 
  \hline
  $\phi_{64}$ &  0.775  & 0.775 \\ \cline{1-3} 
  \end{tabular}
  \end{center}
  \caption{The CA and model performances ${\cal P}_{10^4}^{149}$ of
    $\phi_{8}$, $\phi_{13}$, $\phi_{17}$, $\phi_{18}$, $\phi_{33}$, 
    and $\phi_{64}$. (After Ref. \cite{CrutchfieldEtAl97a}.) Note
	\caS{par}{a} has been referred here as \ca{64}. The CA rule tables
    are given in Table \ref{evhistory-table}.
    \label{SPEResults}
        }
\end{table}

A comparison of the performances of the six CAs just analyzed and their
particle models are given in Table~\ref{SPEResults}. As can be seen,
the agreement is within a few percent for most cases.  In these and
the other cases small discrepancies are due to simplifications
made in the particle models. These include assumptions such as the
particles being zero width and interactions occurring instantaneously.
These error sources are analyzed in depth in 
Ref. \cite{CrutchfieldEtAl97a}.  For $\phi_{18}$ the error is higher,
around $8$\%, due to a long-lived transient domain that is not part of
the particle catalog used for the model.  The main effect of this is
that the condensation time is overestimated on some ICs that generate
this domain. This, in turn, means that the model describes only the
last stages of convergence to the answer configurations, which it gets
correctly and so has a higher performance than $\phi_{18}$. For
$\phi_{33}$ the error is around $4$\%. This appears to be due to
errors in estimates of the distribution of particle types at the
condensation time.

The conclusion is that the particle-level descriptions can be used to
quantitatively predict the computational behavior of CAs and so also
the CA fitnesses and performances in the evolutionary setting. In
particular, the results support the claim that it is these higher-level
structures, embedded in CA configurations, that implement the CA's
computational strategy. More germane to the preceding natural history
analysis, this level of description allows us to understand at a
functional level of structural components the evolutionary process
by which the CAs were produced.

%%%%%%%%%%%%%%%%%% RELATED WORK %%%%%%%%%%%%%%%%%%%%%%%%%

\section{Related Work \label{related-work}}

In Sec.~\ref{computational-task} we discussed some similarities and
differences between this work and other work on distributed parallel
computation. In this section we examine relationships between this
work and other work on computation in cellular automata.

It should be pointed out that \caS{par}{a}'s behavior
(and the behavior of many of the other highest-performance rules) is
very similar to the behavior of the so-called G\'{a}cs-Kurdyumov-Levin
(GKL) CA. This CA was invented not to perform the $\rho_c = 1/2$ task,
but to study reliable computation and phase transitions in
one-dimensional spatially-extended systems \cite{Gacs85a}.  More
extensive work by G\'{a}cs on reliable computation with CAs is
reported in Ref. \cite{Gacs86}. 

The present work and earlier work by our group came out of follow-on research
to Packard's investigation of ``computation at the edge of chaos'' in
cellular automata \cite{Packard88}. Originally Wolfram proposed a
classification of CAs into four behavioral categories \cite{Wolfram83}.
These categories followed the basic classification of dissipative
dynamical systems: fixed point attractors exhibiting equilibrium
behavior, limit cycle attractors exhibiting periodic behavior, chaotic
attractors exhibiting apparently random behavior, and neutrally stable
systems at bifurcations exhibiting long transients. Wolfram suggested
that the latter category was particularly appropriate for implementing
sophisticated (even universal) computation.

Following this with a more quantitative proposal Langton
\cite{Langton90} hypothesized that a CA's $\lambda$---the fraction of
``non-quiescent states'' (here, $1$s) in its look-up table's output
states---was correlated ``generically'' with the CA's computational
capabilities. In particular, he hypothesized that CAs with certain
``critical'' $\lambda$ values, which we denoted $\lambda_c$, would be
more likely than CAs with $\lambda$ values away from $\lambda_c$ to be
able to perform complex computations, or even universal computation.
Packard's goal was to test this hypothesis by using a genetic
algorithm to evolve $(k,r) = (2,3)$ CAs to perform the $\rho_c = 1/2$
task, starting from an initial population chosen from a distribution
that was uniform over $\lambda \in [0,1]$. He found that after 100
generations, the final populations of CAs, when viewed only as
distributions over $\lambda$, tended to cluster close to $\lambda_c$
values. He interpreted this clustering as evidence for the
hypothesized connection between $\lambda_c$ and computational ability.

In Ref. \cite{MitchellEtAl93a} we were able to show, via theoretical
arguments and empirical results, however, that the most successful
CAs for the $\rho_c=1/2$ task must have $\lambda \approx 1/2$. This
value of $\lambda$ is quite different from Packard's quoted $\lambda_c$
values. We argued that Packard's results were due to an artifact in
his particular implementation of the GA. Using more standard versions
and his version of GA search we obtained results that disagreed with
Packard's findings and that were roughly in accord with our theoretical
predictions that high performance CAs were to be found at
$\lambda \approx 1/2$, far from $\lambda_c$, and not in, for example,
Wolfram's fourth CA category. We were also able to explain the
deviations of our results from the theoretical predictions. The current
work came out of the discovery of phenomena, such as embedded-particle
CAs \cite{Crutchfield&Mitchell95a,DasEtAl94a}, that were not found in
Ref. \cite{Packard88}. Moreover, according to Langton the
$\lambda =1/2$ value for our high-performance CAs corresponds to CAs
in Wolfram's chaotic class. The space-time diagrams shown earlier
demonstrate that they are not ``chaotic''; their behavior, in fact,
puts them in the first (fixed-point) category.

Later, other researchers performed their own studies of evolving
cellular automata for the $\rho_c = 1/2$ task.  Sipper and
Ruppin \cite{Sipper97,Sipper&Ruppin96} used a version of the GA to
evolve ``nonuniform CAs''---CA-like architectures in which each cell
uses its own look-up table to determine its state at each time step.
For a lattice of size $N$, the individuals in the GA population are
the $N$ look-up tables making up a nonuniform CA. Sipper and Ruppin used
this framework to evolve $r=1$ nonuniform CAs to perform the $\rho_c =
1/2$ task, as well as other tasks. They reported the discovery of
nonuniform CAs with \perf{10^4}{149} values comparable to that of
\caS{par}{a}.  They did not report \perf{10^4}{N} results for any
other value of $N$ nor did they give statistics on how often
high-performance nonuniform CAs were evolved. Moreover, no structural
analysis of CA space-time behavior or GA population dynamics was
given. Thus, it is unclear how the high fitnesses were obtained,
either dynamically or evolutionarily.

Andre et al. used a genetic programming algorithm to evolve $(k,r) =
(2,3)$ CAs with $N=149$ to perform the $\rho_c=1/2$ task
\cite{AndreEtAl96}. This algorithm discovered particle CAs with higher
\perf{10^4}{149} than that of \caS{par}{a} (e.g., 0.828 versus 0.776).
We obtained the look-up table for one such CA, $\phi_{\tinytextsub{GP}}$
(D.  Andre, personal communication) and found that on larger lattices,
the performance of $\phi_{\tinytextsub{GP}}$ was close to that of
\caS{par}{a} 
(\performance{10^4}{599}{$\phi_{\tinytextsub{GP}}$} $= 0.765$
and \performance{10^4}{999}{$\phi_{\tinytextsub{GP}}$} $= 0.723$;
cf. Table \ref{ca-table}).  
It is not clear whether the improvement in \perf{10^4}{149} was due to
the genetic programming representation CA look-up tables or some other
factor related to increased computational resources. For example,
their runs had a $500$-fold larger population size $M$ and $10$-fold
larger number of ICs over our GA runs. Their runs did, however,
find high-performance CA in average numbers of generations that were
half those in our GA. Thus, the computational resources they used in
their evolutionary search were approximately $2500$ times larger than
in our GA runs.

Paredis \cite{Paredis97} and Juill\'{e} and Pollack
\cite{Juille&Pollack98} experimented with coevolutionary learning
techniques to improve the GA's search efficiency to find embedded
particle CAs for the $\rho_c=1/2$ task. The latter work specifically
rewarded or penalized ICs of particular densities, depending on the
amount of information ICs of those densities provided for
distinguishing fitnesses between CAs in the population. This resulted
in a higher percentage of GA runs in which high-performance
embedded-particle CAs were discovered and in the discovery of
higher-performance CAs than in any of the non-coevolutionary runs. The
highest performance CA discovered had \perf{149}{10^5} $=0.863 \pm
0.001$, \perf{599}{10^5} $=0.822 \pm 0.001$, and \perf{999}{10^5}
$=0.804 \pm 0.001$.  Unfortunately, the performance of this coevolved
CA, although high on small lattices (e.g. $N=149$), decays more
rapidly with lattice size than the GKL rule, which happens to have
lower performance than the coevolved rule on small lattices. This is
appears to be the result of the more complex domains that preclude,
through additional persistent particles, convergence to the answer
configurations, $0^N$ or $1^N$.  Compared to the coevolved CAs, the
GKL CA is one of the CAs that maintains high performance on larger
lattices.

Our own work has been extended to other tasks, most thoroughly to a
global synchronization task for which we have performed similar
analyses to those given in this paper \cite{DasEtAl95a}.

Our notion of computation via particles and particle-interactions
derives from that introduced by the computational mechanics framework
\cite{Crut92c,Hanson93,Hans90a} and so differs considerably from the
notions used in most other work on designing CAs for computation.  For
example, propagating particle-like signals were used in the solution
to the Firing Squad Synchronization Problem
\cite{Mazoyer88,Moore64,Waksman66}, in Smith's work on CAs for
parallel formal-language recognition \cite{Smith72}, and in Mazoyer's
work on computation in one-dimensional CAs \cite{Mazoyer96}.  However,
in all these cases, the particles and their interactions were designed
by hand to be the explicit behavior of the CA. That is, the particles
are explicitly coded in each cell's local state and their dynamics and
their interactions are coded directly into the CA lookup table.
Typically, their
interactions were effected by a relatively large number of states per
site.  Steiglitz, Kamal, and Watson's carry-ripple adder
\cite{SteiglitzEtAl88} and the universal computer constructed in the
Game of Life \cite{BerlekampEtAl82} both used binary-state signals
consisting of propagating periodic patterns. But, again, the particles
were explicitly designed to ride on top of a quiescent background and
their interaction properties were carefully hand coded.  In Squier and
Steiglitz's ``particle machine'' \cite{Squier&Steiglitz94} and in
Jakubowski, Steiglitz, and Squier's ``soliton machine''
\cite{JakubowskiEtAl96}, particles are the primitive states of the CA
cells. Moreover, their interaction properties are explicitly given by
the CA rule table.  These machines are essentially kinds of lattice
gas automata \cite{Doolen90} that operate on ``particles''
directly. (Other work on arithmetic in cellular
automata has been done by Sheth, Nag, and Hellwarth \cite{ShethEtAl91}
and Clementi, De Biase, and Massini \cite{ClementiEtAl94}, among
others.)

In contrast to these, particles in our system are embedded
as walls between regular domains. They are often apparent only after
those domains have been discovered and filtered out. Their structures
and interaction properties are emergent properties of the patterns
formed by the CAs. Notably, although each cell has only two possible
states, the structures of embedded particles are spatially and
temporally extended, and so are more complex than atomic or simple
periodic structures. Typically, these structures can extend over
spatial scales larger than the CA radius. For example, the background
domain of the elementary CA (ECA 110) shown in \fig{eca110} has a
temporal periodicity of $7$ time steps and a spatial periodicity of
$14$ sites, markedly larger than the $r=1$ nearest-neighbor coupling.

%%%%%%%%%%%%%%%%%% CONCLUSION %%%%%%%%%%%%%%%%%%%%%%%%%

\section{Conclusion}

Our philosophy is to view CAs as systems that naturally form patterns
(such as regular domains) and to view the GA as taking advantage---via
selection and genetic variation---of these pattern-forming
propensities so as to shape them to perform desired computations.
Within this framework, we attempt to understand the behavior of the
resulting CAs by applying tools, such as the computational mechanics
framework, formulated for analyzing pattern-forming systems. The
result gives us a high-level description of the computationally
relevant parts of the system's behavior.  In doing so, we begin to
answer Wolfram's last problem from ``Twenty problems in the theory of
cellular automata'' (Wolfram, 1985): ``What higher-level descriptions
of information processing in cellular automata can be given?'' We
believe that this framework will be a basis for the ``radically new
approach'' that Wolfram claimed will be required for understanding
and designing sophisticated computation in CAs and other decentralized
spatially extended systems.

Our analysis showed that there are three levels of information
processing occurring during iterations of the evolved high-performance
CAs. The first was the type of information storage and transmission
effected by the particles and the type of ``logical'' operations
implemented by the particle interactions. The second, higher level
comprised the geometric subroutines that implemented intermediate-scale
computations. We analyzed in detail two of these that were important to
the size competition between regions of low and high density. We also
showed how variations in the particles led to several types of error
at this level. The third and final level is that of the global
computation over the entire lattice up to the answer time. This is
the level at which fitness is conferred on the CAs.

We analyzed in some detail the natural history that led to the
emergence of such computationally sophisticated CAs. The evolutionary
epochs typically proceed in a set sequence, with earlier epochs
setting the (necessary) context for the later, higher performance
ones. Often the jumps to higher epochs were facilitated by
exaptations---changes in adaptively neutral traits appearing in much
earlier generations.

There are a number of fruitful directions for future work.  The first
is to extend the lessons learned here to more general evolutionary
search algorithms and pattern forming dynamical systems.  The problem
of choosing a genetic representation of dynamical systems that helps,
or at least does not hinder, the search will play an important role in
addressing this. The evolution of CAs that operate on two-dimensional
images rather than one-dimensional strings will also help address this
issue and also open up application areas, such as iterative nonlinear
image processing \cite{Crut88c}.

We also need to develop substantially better analytical descriptions
of the search's population dynamics and of how the intrinsic
structures in CAs interact with that dynamics. Although the evolution
of CAs is a very simplified problem from the biological perspective,
the evolutionary time scale of the population dynamics and the
development time scale of the CAs result in a two-time-scale
stochastic dynamical system that is difficult to analytically predict.
Such predictions, say of how to set the mutation rate or population
size for effective search, are centrally important both for basic
understanding of evolutionary mechanisms and for practical
applications. Progress on quantitatively predicting the population
dynamics occurring during epochal evolution has been made
\cite{Nimw98a,Nimw97a}. The adaptation of the ``statistical dynamics''
approach introduced there to the evolution of CAs will be an
important, but difficult, step toward understanding complicated
genotype-to-phenotype maps. The latter is highly relevant for using
such search methods on complex problems.

Another quantitative direction is the estimation of computational
performance of distributed systems based on higher-level descriptions.
The results, reported here and described in detail in Refs.
\cite{CrutchfieldEtAl97a} and \cite{HordijkEtAl98}, on predicting CA
computational performance are encouraging.  Constructing a more
accurate model along with a quantitative analytical model of
higher-level computation in CAs will help us understand how much the
embedded CA structures contribute individually to overall fitness. And
this, in turn, will allow us to monitor the evolutionary mechanisms that
lead to the emergence of collective computation in coordinated groups of
functional units.

% how general?  how general to this problem, 1D CAs, CAs in general
% 
% Significance.  Density classification -- not per se interesting (although
% perhaps it is) but shows how CAs can do collective computation.  
% 
% Constellations of particles, predicting epochs.  
% 
% Take off times.
% 
% Discuss representation issues
% 
% Talk about future work -- 2D, coevolution
% 
% Mention GA-mechanisms paper. 

% Need to expand computation theory, develop vocabulary for collective
% computation -- this is one direction.

\section*{Acknowledgments}

The authors thank Wim Hordijk for calculating the CA particle-model
performances and Silas Alben for calculating the coevolved CA
performances. They also thank Jim Hanson, Wim Hordijk, Cris Moore,
Erik van Nimwegen, and Cosma Shalizi for helpful discussions. This
work was supported at UC Berkeley by ONR grant N00014-95-1-0524 and at
the Santa Fe Institute by ONR grant N00014-95-1-0975, Sandia National
Laboratory contract AU-4978, and National Science Foundation grant
IRI-9705853.

\appendix
\section{Domain Filter \label{PhiParATransducer}}

In this appendix we describe the properties and construction of
\caS{par}{a}'s domain-recognizing and filtering transducer.

\begin{figure}
\centerline{\psfig{figure=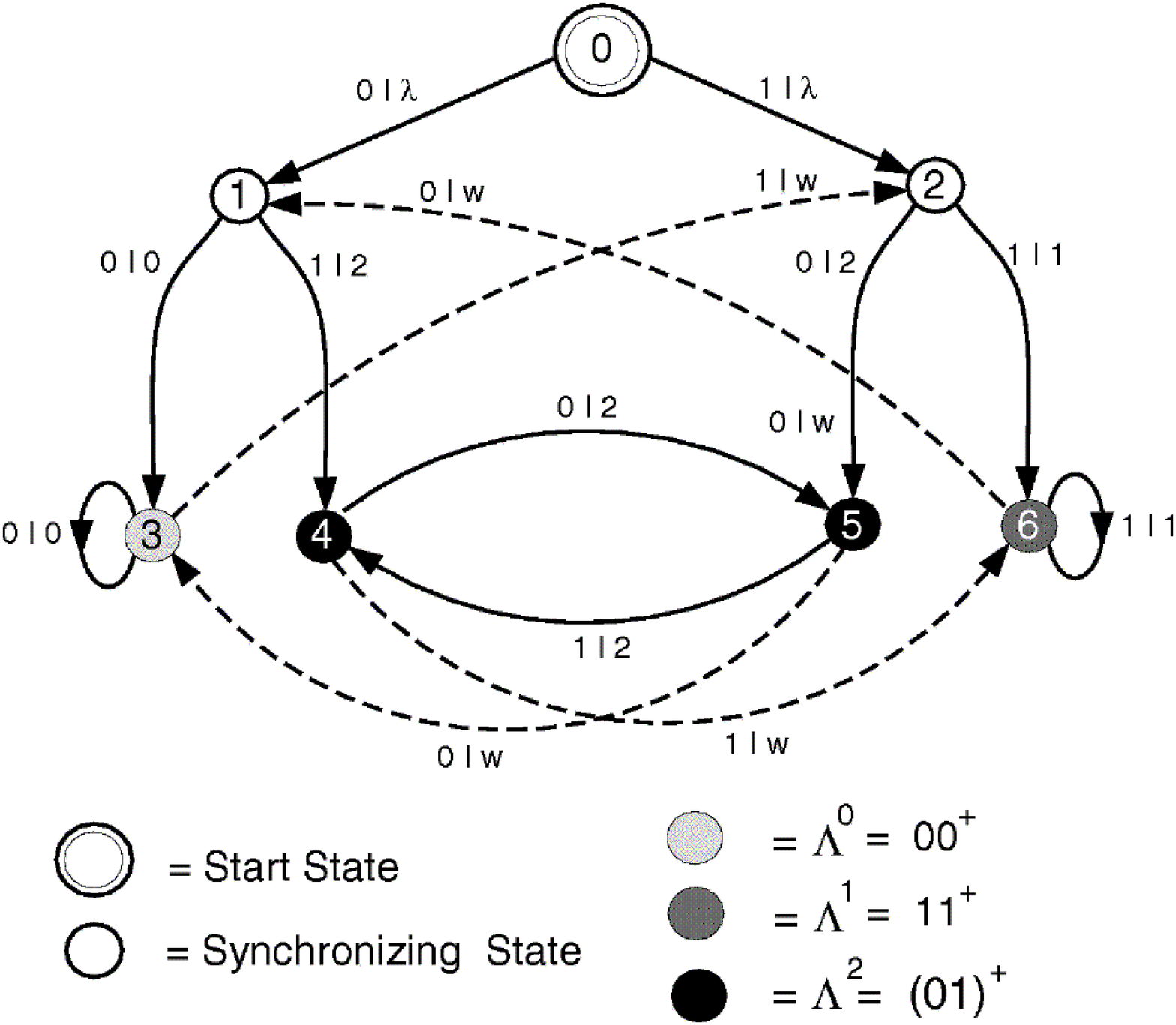,height=5in}}
\caption{
  \caS{par}{a}'s domain-recognizing and filtering transducer.
  The edge labels $s|t$ indicate that the transition is to
  be taken on reading configuration site value $s \in \{0,1\}$
  and then outputting structural label
  $t \in \{ \lambda, 0, 1, 2, w \}$.
  \label{PhiParATransducerFig}
  }
\end{figure}

The transducer, shown in \fig{PhiParATransducerFig}, reads in binary
CA configurations and outputs strings of the same length,
the lattice size $N$, in the domain-wall alphabet
$\{ \lambda, 0, 1, 2, w \}$. In this alphabet $\lambda$ indicates that
the transducer has not yet ``synchronized'' (see below) to the domain or wall
structures in the configuration, $\{ 0,1,2\}$ label each of the three
domains, respectively, and $w$ indicates a wall between domains. In
the filtered space-time diagrams $w$ is mapped to black and all other
output symbols map to white.

Briefly, \caS{par}{a}'s domain-wall transducer is constructed as
follows. \caS{par}{a} has three domains, each of which can be
described by simple finite-state machines. These machines form the
recurrent states of the transducer. When the transducer first begins
to read in the configuration, it may take several steps to disambiguate
the site values and identify the appropriate domain in which they are
participating. Working through the transitions and transient states
that lead to the recurrent (domain) states determines the transitions
from the start state. When the transducer is reading site values
consistent with one of these domains, but then encounters site values
that are not consistent with it (e.g. values indicating walls), then
some number of additional site values must be read in to determine the
domain type into which the transducer has moved. Such transitions
determine the transducer's domain-to-domain transitions. 

Note, that due to the steps required to initially read in a sufficient
number of site values to recognize the domains and walls, a process that
we call synchronization, the transducer may have to read some portion of
the configuration that it has already read, as it wraps around due
to the lattice's periodic boundary conditions. This takes at most
one additional pass over the configuration. 

The general construction procedure for domain-wall transducers is
given in Ref. \cite{Crut93a}.

%%%%%%%%%%%%%%%%%%% PAR_A TRANSDUCER %%%%%%%%%%%%%%%%%%%
%%%%%%%%%%%%%%%%%%%%%%%%%%%%%%%%%%%%%%%%%%%%%%%%%%%%%%%%

\bibliography{evca}
\bibliographystyle{plain}

\end{document}